\documentclass[traditabstract]{aa}

\usepackage{multirow}

\usepackage{caption}
\usepackage{subcaption}
\usepackage{soul}

\usepackage{amsmath}
\usepackage{graphicx}

\usepackage{txfonts}
\usepackage{comment}
\usepackage{natbib}
\bibpunct{(}{)}{;}{a}{}{,}
\usepackage{float}
\usepackage{calc}
\usepackage{textcomp}
\usepackage{placeins}
\usepackage{color}
\usepackage{rotating} 
\usepackage{lscape}
\usepackage{url}
\usepackage[colorlinks=true,linkcolor=blue,citecolor=blue]{hyperref}
\usepackage[all]{draftcopy}

\usepackage{longtable}
\usepackage{array}
\usepackage{threeparttable}
\usepackage{lscape}

\usepackage{enumerate}
\DeclareTextSymbol{\degre}{OT1}{23}

\renewcommand{\arraystretch}{1.2}
\newcounter{savedfootnote}

\def \cigale{{{\sc cigale}}}

\begin{document} 

\title{Stochastic star formation activity of galaxies within the first billion years probed by JWST}

\author{
C.~Carvajal-Bohorquez\inst{1}\fnmsep\thanks{\email{cristian.carvajal@lam.fr}},
L.~Ciesla\inst{1},
N.~Laporte\inst{1},
M.~Boquien\inst{2},
V.~Buat\inst{1},
O.~Ilbert\inst{1},
G.~Aufort\inst{3},
M.~Shuntov\inst{4,5,6},
C.~Witten\inst{6},
P.~A.~Oesch\inst{6,4,5}
and A.~Covelo-Paz\inst{6}
}

\institute{	
Aix Marseille Univ, CNRS, CNES, LAM, Marseille, France
\and
Universit\'e Côte d'Azur, Observatoire de la Côte d'Azur, CNRS, Laboratoire Lagrange, 06000, Nice, France
\and
Institut d'Astrophysique de Paris, UMR 7095, CNRS, Sorbonne Université, 98 bis Boulevard Arago, F-75014, Paris, France
\and
Cosmic Dawn Center (DAWN), Denmark
\and
Niels Bohr Institute, University of Copenhagen, Jagtvej 128, 2200 Copenhagen, Denmark
\and
Department of Astronomy, University of Geneva, Chemin Pegasi 51, 1290, Versoix, Switzerland
}	

   \date{Received ; accepted }

\abstract
{
Early observations with the \textit{James Webb Space Telescope} have highlighted the excess of UV-bright galaxies at $z>10$, with a derived UV luminosity function (UVLF) that exhibits a softer evolution in redshift than expected. This unexpected trend may result from several proposed mechanisms, including a high star formation efficiency (SFE) or a bursty star formation history (SFH).  
In this work, we aim to characterize the burstiness level of high redshift galaxy SFHs and its evolution.
We implemented a stochastic SFH module in \cigale\ using power spectrum densities, to estimate the burstiness level of star formation in galaxies at $6<z<12$.
We find that SFHs with a high level of stochasticity better reproduce the Spectral Energy Distributions of $z>6$ galaxies, while smoother assumptions introduce biases when applied to galaxies with bursty star formation activity.
The assumed stochasticity level of the SFH also affects the constraints on galaxies' physical properties, producing a strong and tight relation between the star formation rate (SFR) and stellar mass in the case of a smooth SFH, down to a weak relation at $z\geq7$ for an SFH with a high level of stochasticity.
Successively assuming different levels of burstiness, we determined the best-suited SFH for each $6<z<12$ galaxy in the JADES sample from a Bayes factor analysis. Galaxies are classified according to their level of burstiness, and the corresponding physical properties are associated with them. 
For massive galaxies (8.8 $< \log M_\star/M_\odot <$ 9.5), the fraction of bursty galaxies increases from 0.38$\pm$0.08 to 0.77$\pm$0.2 at $z \sim 6$ and $z \sim 12$, respectively. At all redshifts, only $<20$\% of low-mass galaxies are classified as bursty; although, this estimate is uncertain because their faintness leads to a low signal-to-noise ratio.
For bursty galaxies, the $\rm \log_{10}(SFR_{10}/SFR_{100})$ ratio, another indicator of bursty star formation, does not evolve with redshift, but the fraction of galaxies with a high $\rm \log_{10}(SFR_{10}/SFR_{100})$ slightly increases from 0.28$\pm$0.06 to 0.38$\pm$0.11 between $z\sim6$ and $z\sim9$.
We include additional constraints from observations on $\sigma_{UV}$, the dispersion of the UV magnitude distribution, and SFE, finding a maximum of 0.72$\pm$0.02\,mag and 0.06$\pm$0.01 for $\sigma_{UV}$ and SFE, respectively. This confirms that neither alone is responsible for the weak evolution of the UVLF at $z>10$.
Our results add further evidence that a combination with other mechanisms is likely responsible for the high-$z$ UVLF.
The stochastic SFH module is public as part of \texttt{CIGALE} version 2025.1.
}

\keywords{Galaxies: evolution; Galaxies: high-redshift; Galaxies: star formation}

   \authorrunning{Carvajal-Bohorquez et al.}
   \titlerunning{Stochastic star formation in early galaxies}
   
   \maketitle

\section{Introduction}\label{sec: intro}

The physical process that governs the evolution of galaxies and shapes their properties,
such as their mass assembly history and the underlying star formation process, are the cornerstone to understand the structures that we observe at the present. Secular star formation dominates in most of the galaxies at $z\lesssim5$, leading to the well-established main sequence (MS) of star-forming galaxies \citep{Noeske07_SFseq,Daddi07,Elbaz07}. The \textit{Hubble Space Telescope} (HST) pushed the constraint on galaxy evolution models by revealing the evolution of the ultraviolet luminosity function (UVLF) at $z\sim8$ \citep{Bouwens15,Finkelstein15}. In the pre-JWST era, our understanding of the universe at $z > 8$ showed limited constraints on the UVLF and its evolution toward higher redshifts \citep{Oesch18,Bouwens19,Finkelstein22a,Bagley24}. The revolutionary sensitivity and resolution of JWST in the near-infrared have opened a new cosmic window, facilitating the robust detection and characterization of galaxies at $z>8$. 

The first JWST studies of the high redshift Universe revealed an unexpectedly numerous population of bright galaxies at $z > 10$ \citep[e.g.,][]{Finkelstein22b, Naidu22b, Castellano22, Donnan23a}. These early results suggested a UVLF with a surprisingly mild evolution during the first 500 million years after the Big Bang \citep{Harikane23a, Donnan23, Adams24, Finkelstein24, Robertson24}. This result contrasts with empirical extrapolations based on pre-JWST data \citep[e.g.,][]{Bowler20,Bouwens21,FinkelsteinBagley22} at $z\sim11$ and $z\sim14$ \citep{Finkelstein24}. Different scenarios have been proposed to account for this overabundance with physical processes that enhance the UV emission via the following: a top-heavy initial mass function \citep{Cueto24, Hutter25}, reduced dust attenuation \citep{Ferrara23}, super massive black holes \citep{Matteri25}, increased stochasticity in the star formation (SF) \citep{Mason23, PallottiniFerrara23, Shen23, Gelli24, Ciesla24, Sun24}, or high star formation efficiency \citep{Munoz23, Shuntov25}. 

At low redshift, the brightest galaxies typically reside in the most massive halos \citep[e.g.,][]{Behroozi10, Behroozi13, Allen19}. At high redshift, stochasticity becomes more significant, directly impacting the UVLF \citep{Shen23}. This stochasticity can be characterized by the dispersion in the UV magnitude-halo mass ($M_{\rm UV}-{\rm M_h}$) relation, $\rm \sigma_{UV}$ \citep{Mashian16, Ren19, Shen23, Mason23, Gelli24,Shen24,Shuntov25}.
Consequently, the brightest galaxies do not need to reside within the most massive halos, but can instead reside within the more abundant, lower-mass halos. This dispersion would arise as a natural result of bursty star formation histories (SFHs), within which galaxies residing in low-mass halos are more prone to experience \citep{Sun24, Gelli24}. This behavior has been identified in simulations \citep{Dayal13, KimmCen14, Faisst19, Emami19, Wilkins23, Sun23, PallottiniFerrara23,Basu25}, and observationally from the scatter of the MS \citep{Cole25,Clarke24} and signatures in the recent SFH \citep{Ciesla24, Endsley24b}.

JWST-NIRSpec has revealed that high redshift galaxies exhibit a broad range of star formation activity, highlighting a population of galaxies in a temporarily suppressed star formation phase, predicted by simulations \citep[e.g.,][]{Wyithe14}, identified in the literature as ``lull" or ``dormant,'' when presenting weak emission lines, and ``mini-quenched,'' when presenting no emission lines \citep[e.g.,][]{Strait23,Faisst24,Looser25,Baker25,CoveloPaz25,Witten25b}. These galaxies usually undergo an intense burst of SF followed by a rapid quenching. A good example is the mini-quenched galaxy at $z\sim7$ reported in \cite{Looser24a}, which has a stellar mass of $\rm \log(M_\star/M_\odot)=8.7$ and shows a suppressed star formation phase lasting 10-20\,Myr before the epoch of observation \citep{Dome24,McClymont25,Witten25a,CoveloPaz25}. Low-mass galaxies in this range ($\rm \log M_\star/M_\odot\sim8$) are more sensitive to feedback mechanisms due to their shallow gravitational potential \citep{Hopkins23, Gelli25}, which can produce a temporary or even permanent quiescence. In the \textsc{serra} simulations, \cite{Gelli25} \citep[see also][]{Gelli23} found that quiescent systems dominate the population at $\log(M_\star/M_\odot)<8$, exhibiting bursty SFHs. This transitional stellar mass, where galaxies are dominated by stochasticity rather than continuous SF, has also been identified in \textsc{FIRE-2} zoom-in simulations via the MS relation, where galaxies with $\log(M_\star/M_\odot)<8$ show a large scatter of 0.6\,dex \citep{Ma18}. Studies of local analogs of primordial galaxies show a similarly broad distribution of H$\alpha$-to-UV luminosity ratios in systems with $\log(M_\star/M_\odot)<8$ \citep{Emami19}. Using nonparametric SFHs, \cite{Ciesla24} found a transition between bursty and smooth SFHs at $\log(M_\star/M_\odot)<8.6$ in galaxies at $6 < z < 7$. 

The diversity of star formation activity revealed by JWST observations of high redshift galaxies points out the importance of adopting more flexible approaches to modeling their SFHs. Although nonparametric SFHs do not assume a predefined shape, they may not capture the rapid and irregular variations suggested by observations. Alternatively, \cite{CaplarTacchella19} proposed a formulation that incorporates the stochasticity in a time-dependence model through a power spectrum density (PSD). This approach has been used in the literature to model observations of early galaxies \citep[e.g.,][]{Faisst24}, and in theoretical modeling to reproduce the UVLF \citep[e.g.,][]{KravtsovBelokurov24, Sun24}. In this work, we implement this stochastic SFH approach in {\cigale} to characterize the burstiness level of high redshift galaxies ($6<z<12$) and its evolution. We present and validate our SFH framework in Sect.~\ref{sec: stochacti_SFH_cigale}. The sample is presented in Sect.~\ref{sec: sample}. In Sect. \ref{sec: sed_modeling}, we perform Spectral Energy Distribution (SED) fitting implementing the stochastic SFH approach and, we discuss its impact on the MS and the derived physical properties of the galaxies. Base on the SED modeling, we investigate whether the galaxies in our sample exhibit bursty SFHs in Sect.~\ref{sec: galaxy_classification}. We discuss the evolution of the galaxies' burstiness via $\rm \log SFR_{10}/SFR_{100}$ and $\sigma_{UV}$ as tracer of stochastic star formation in Sect.~\ref{sec: evolution_gal_burstiness} and \ref{sec: sigma_uv_redshift}, respectively. Finally, we summarize our results in Sect.~\ref{sec: conclutions}. Throughout this paper, we use a \cite{Chabrier03} initial mass function, we adopt a standard $\Lambda$CDM cosmology with $H_0 = 67.66\,\mathrm{km\,s^{-1}\,Mpc^{-1}}$, $\Omega_{\mathrm{m},0} = 0.3$, $\Omega_{\mathrm{b},0} = 0.048$, and $\Omega_{\Lambda,0} = 0.68$ \citep[Planck18 in][]{astropy:2022}; and magnitudes are in the AB system \citep{Oke&Gunn}.

\section{Stochastic star formation histories in \cigale}\label{sec: stochacti_SFH_cigale}

We used the \cigale\footnote{\url{https://cigale.lam.fr/}} code \citep{Boquien19} to model the SED of galaxies, estimate their physical properties, and reconstruct their SFH. This versatile code combines multiple modules to model the different physical components used to build SEDs. In this work, we modeled the SED using single stellar populations using \cite{BruzualCharlot03} models, adding nebular emission \citep{Inoue11}, and modeling the dust attenuation using a \cite{Calzetti00} law. Apart from the SFH assumption that we describe below, we used the input set of parameters provided and tested by \cite{Ciesla24} on early galaxies. However, we limit the range of attenuation, as probed by E(B-V)$_{\rm lines}$, to 0-0.2 following the results of \cite{Ciesla25}. Using ALMA constraints obtained from stacking $z>6$ galaxies, they found that this range of attenuation is well-suited for galaxies at these redshifts, especially at $z>9$. This range is also consistent with the results of \cite{Saxena24} and \cite{Donnan25}, using spectroscopy.

In SED modeling, one of the key assumptions is the SFH. In \cigale\, different analytical shapes are implemented, such as exponentially declining or rising SFH and $\tau$-delayed models. Recently, nonparametric models have also been implemented and tested in \cigale\ \citep{Ciesla23,Ciesla24,ArangoToro23}. This approach helps to mitigate the known biases resulting from the use of simple analytical functions \citep[e.g.,][]{Buat14,Ciesla15,Ciesla17,Carnall19,Iyer17,Iyer19,Leja19a,Lower20,Leja22,Wang25}.
The nonparametric approach does not assume any analytical function, but rather a number of time intervals where the star formation rate (SFR) is constant. These time bins are linked between them using a prior that weights against a sharp transition. The length of the time bins and the priors used might not be suitable to model the SFH of early galaxies for which increasing evidence shows that they are bursty \citep[e.g.,][]{Ciesla24,Endsley24b,Cole25,Kokorev25}. To adapt to the burstiness, \cite{Ciesla24} proposed to use a flat prior instead to give enough freedom to reproduce rapid and intense changes in star formation activity \citep[see also for a stochastic prior,][]{Wan24}.
In this work, we aim to go a step further by building stochastic SFHs that would allow for increased flexibility in the levels of burstiness.

To this aim, we propose to model the SFH with two components: a smooth one that evolves over the time following a parametric approach ($\tau$-delayed model) and a bursty one (see Sect.~\ref{sec: stochastic_module}) that captures the stochastic processes on short time scales. Then, the resulting SFH is the combination of the following two components:
\begin{equation}
    \text{SFR }= \text{SFR}_{\text{smooth}} \times \text{SFR}_{\text{stochastic}}.
    \label{eq: sum_sfh_components}
\end{equation}
The stochastic term is normalized to one, so in the absence of burstiness the SFH reduces to the smooth component, which serves as the baseline, while in bursty scenarios $\rm SFR_{stochastic}$ modulates the overall star formation activity. We develop and test this approach in the following sections. 

\subsection{{\sc{StochasticSFH}} module}\label{sec: stochastic_module}

To model the peaks and troughs of stochastic star formation, we follow the approach described in \cite{CaplarTacchella19}, where the SFR fluctuations are treated as stationary stochastic processes, defined by a power spectral density. The PSD quantifies the distribution of power as a function of frequency, providing a measure of the contribution of different timescales to the overall signal. For instance, white noise is a stochastic process with flat power distribution where all frequencies contribute equally, therefore there is no temporal correlation. Realistic SFH exhibits correlated fluctuations at short time scales, to capture this, we used a broken power-law PSD,
\begin{equation}
\operatorname{PSD}(f)=\frac{\sigma}{1+\left(\tau_{\text {break }} f\right)^\alpha},
\end{equation}
\noindent where $\sigma$ characterizes the amplitude of SFR long term variability or the PSD normalization (burstiness level), $\alpha$ is the slope of the power-law at high frequencies, $\tau_{\text{break}}$ - the correlation time - characterizes the timescale where the fluctuations are no longer correlated, and $f$ is the frequency. With this model, we assume that the stochastic process is stationary, therefore, its mean, variance, and autocorrelation function are not time dependent \citep[for further details, see Sect.~2 in][]{CaplarTacchella19}. 
Physically, the fluctuations modeled by this approach could represent variations in star formation activity due to feedback-driven outflows, mergers, or variations in gas mass accretion rate which can happen on timescales of one to a few hundred Myr \citep{Tacchella20}. 

Practically, we create a new SFH module within \cigale\ where we first define the PSD by providing as input parameters $\alpha$, $\tau_{break}$, and $\sigma$. Then, we derive the time series from the PSD, which is in the frequency domain, by using the algorithm of \cite{Timmerkoning95}. It is implemented in the \texttt{DELightcurveSimulation} Python package by \cite{Connolly16} and already used for the same purpose by \cite{Faisst24}. This method applies an inverse Fourier Transform on the provided PSD to get the time evolution of the SFH. We set the same minimum time sampling of the SFH to 1\,Myr, the default value in \cigale. 
This implies that the SFR is constant over 1\,Myr, and therefore sets the temporal resolution of the PSD. Furthermore, this fixes a lower limit on $\tau_{\text{break}}$, because by construction the SFR is perfectly correlated at $ \leq 1$\,Myr. 

The smooth component of the SFH is the classical $\tau$-delayed model parameterized as in the \texttt{sfhdelayed} module of \cigale,
\begin{equation}
\operatorname{SFR }_{\text{smooth}}(t)=\frac{t}{\tau_{\text{main}}^2} \times \text{exp}(-t/\tau_{\text{main}}),
\end{equation}
\noindent where $\tau_{\text{main}}$ is the e-folding time of the exponential.\\

\begin{figure*}[!h]
\includegraphics[width=0.32\textwidth]{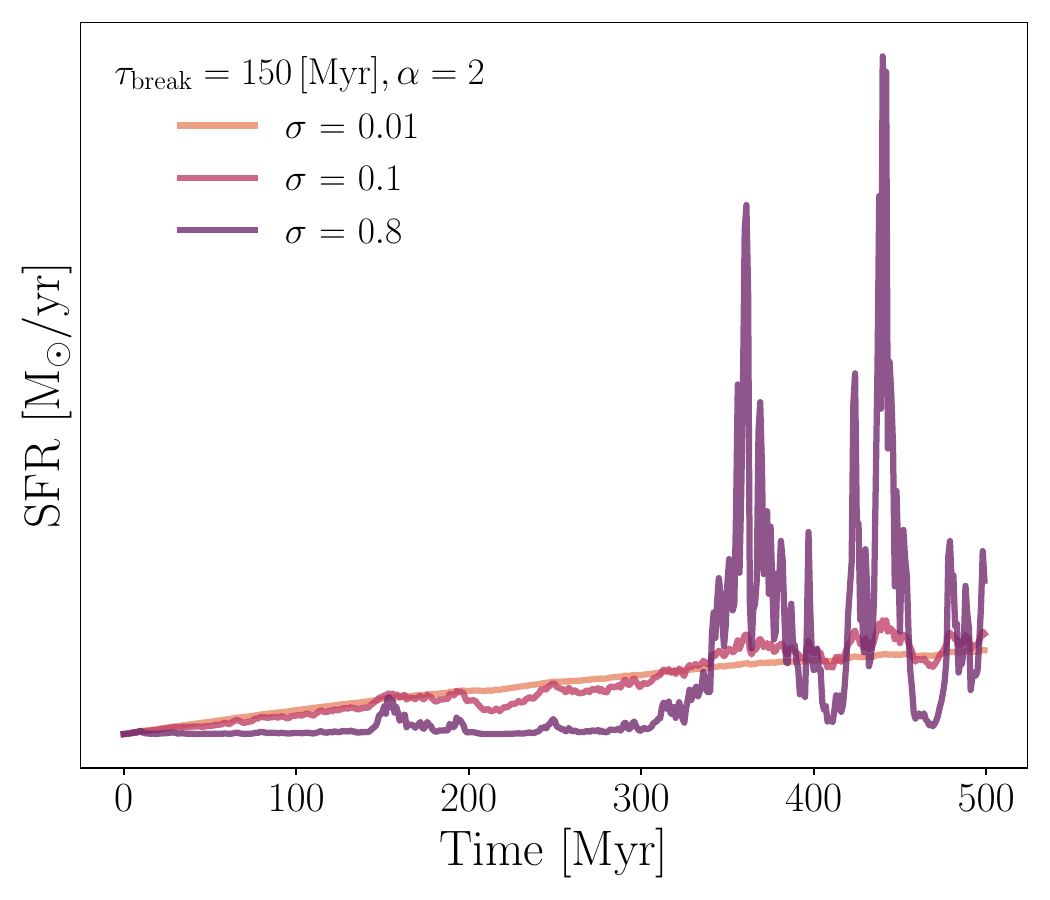}
\includegraphics[width=0.32\textwidth]{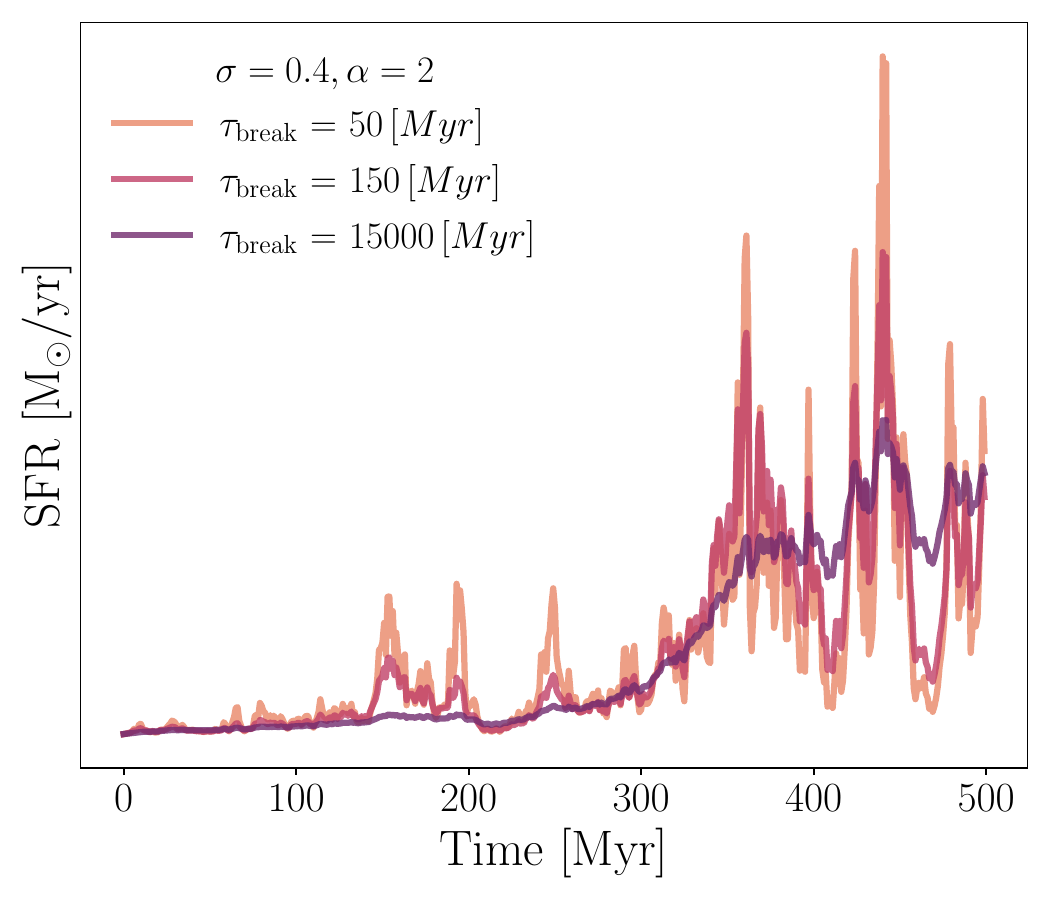}
\includegraphics[width=0.32\textwidth]{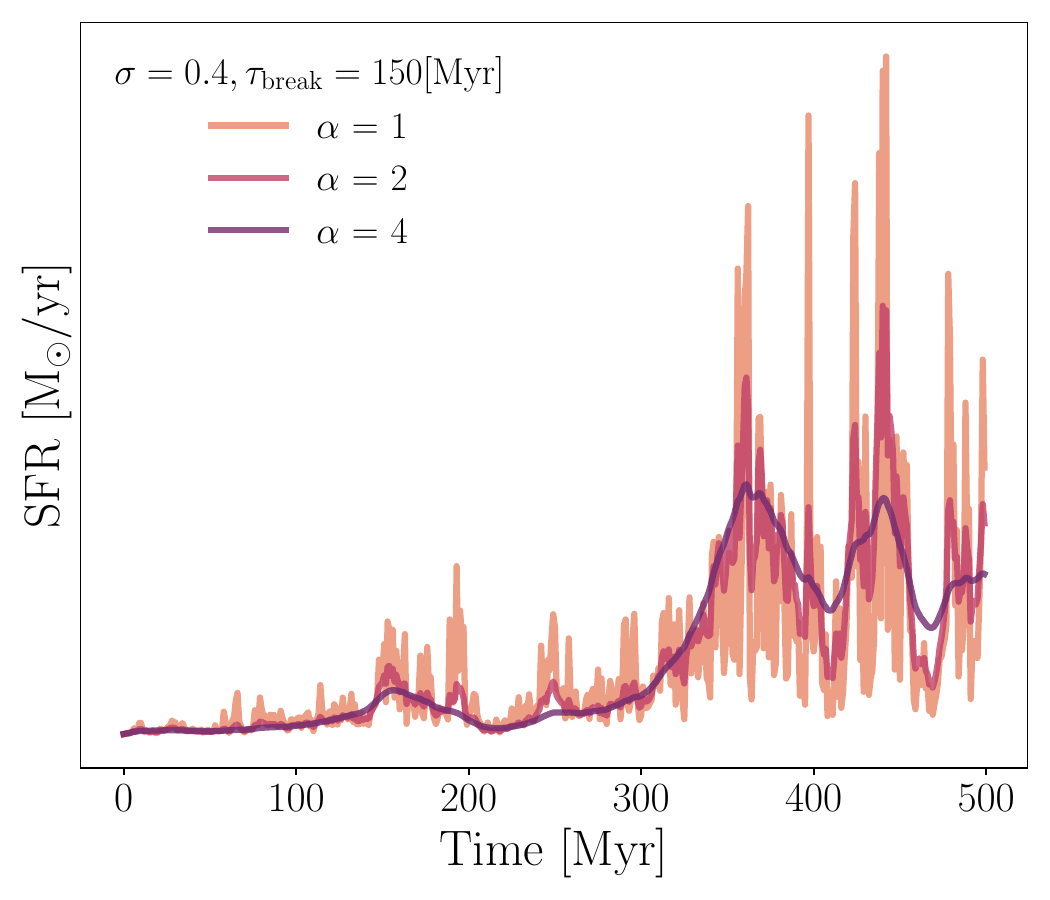}
\caption{Stochastic star formation histories ($\text{SFR}_{\text{smooth}} \times \text{SFR}_{\text{stochastic}}$) for different combinations of parameters generated with the {\sc{StochasticSFH}} module. Each panel shows the impact on the SFH of one specific parameter: $\sigma$ (left panel), $\rm \tau_{break}$ (middle panel), and $\alpha$ (right panel), in all cases the age and $\rm \tau_{main}$ are fix to 500\,Myr and 1000 \,Mys, respectively.}
\label{fig: SSFH}
\end{figure*}

Therefore, we have four free parameters, three ($\sigma$, $\alpha$ and $\tau_{\text{break}}$) from the stochastic component and one ($\tau_{\text{main}}$) from the smooth component, in addition to the length of the SFH that represents the age of the galaxy\footnote{In \cigale, the age of a galaxy is the time since the birth of its first star.}. For each combination of parameters, the module draws a given number of SFH (N$_{\rm SFH}$) using a random number generator. In Fig.~\ref{fig: SSFH}, we show examples of the modeled SFH for different parameter combinations, while fixing the age and $\rm \tau_{main}$ to 500 and 1000\,Myr, respectively. For an SFH dominated by the smooth component ($\sigma < 0.1$), the amplitudes of the fluctuations are small. The shape of the PSD characterizes the SFR fluctuations; small correlation times lead to a short burst-dominated SFH, while long times produce less variation. The power law index $\alpha$ produces a similar effect, increasing the slope leads to a smoother SFH with less fluctuations, as it puts more power into the long-term variations. The stochastic SFH module is public\footnote{\url{https://cigale.lam.fr/2025/10/06/version-2025-1/}.} as part of \texttt{CIGALE} version 2025.1.

Similar descriptions have also been used in the literature to model the SFH of galaxies at high redshift in physically-motivated approaches such as in \cite{Sun24} and \cite{KravtsovBelokurov24}, although in those studies the first term is associated with the mass accretion rate onto dark matter halos, so it is a mass-dependent function.
Recently, \cite{Faisst24} used this approach to model the fluctuations of galaxy SFRs around the MS at a given stellar mass and redshift.
Following their study, we fix $\alpha$ to 2 and $\tau_{break}$ to 150 Myr as these parameters are found difficult to constrain from the observations \citep[e.g.,][]{CaplarTacchella19}. 
We note that these values do not prevent from modeling galaxies in lulling or temporary quenching phases, as shown by \cite{Faisst24}. This leaves two free parameters, $\sigma$ and $\rm \tau_{main}$, in addition to the galaxy age.

\subsection{Test of the {\sc{StochasticSFH}} module}\label{sec: module_test}

We test the ability of the {\sc{StochasticSFH}} module, combined with spectral population models, nebular models, and dust attenuation, to reproduce JWST observations of galaxies at $z\geq6$. In this test, we focus on constraining only the amplitude of the stochastic fluctuations ($\sigma$), since we fixed the other parameters of the {\sc{StochasticSFH}} module. However $\rm \tau_{main}$ is free to provide variations in the smooth component of the SFH.

\begin{table*}[!htbp]
   \centering
   \caption{\cigale\ input parameters used to build the mocks SEDs and to fit the spectrum of galaxies at $z\geq6$.}
   \begin{tabular}{l cl}
   \hline\hline
    Parameter & Value & Definition\\
   \hline
   \multicolumn{3}{c}{Stochastic SFH -- \sc{StochasticSFH}}\\[1mm]  
   $age$ (Myr) & 300, 400, 500, 750 &  For the mock catalog\\
    & 300, 500, 750, if $6<z<7$ & Age of the galaxy \\ 
    & 250, 350, 500, if $7<z<9$ & since first star formation\\ 
    & 150, 225, 300, if $9<z<12$ & \\
   $\tau_{\text{main}}$ (Myr) & 200, 300, 1000 & Time when smooth component peaks\\ [1mm] 
   $\tau_{\text{break}}$ (Myr) & 150 & Timescale where the fluctuations are no longer correlated \\[1mm]
   $\sigma$ & 0.01,0.1,0.2,0.4,0.6,0.8,1.0& One $\sigma$ per run - burstiness level \\[1mm]
   $\alpha$ & 2 & PSD slope \\[1mm]
   $N_{\rm SFH}$ & 500 & \# of SFH per $age$ value \\[1mm]
   & 800 & \\
   \hline
   \multicolumn{3}{c}{Stellar population -- \sc{bc03}}\\[1mm]  
   metallicity      &  0.004, 0.02     & \\   
   \hline
   \multicolumn{3}{c}{Emission lines -- \sc{nebular}}\\[1mm]  
   $\log$U      &  -3, -2     & Ionization parameter\\
   $f_{\rm esc}$      &  0.0, 0.1     & Lyman continuum photons escape fraction\\
   $f_{\rm dust}$      &  0.0, 0.1     & Lyman continuum photons fraction absorbed by dust\\
   \hline
   \multicolumn{3}{c}{Dust attenuation -- \sc{dustatt\_modified\_starburst}}\\[1mm]  
   E(B-V) lines      &  [0.0, 0.2]  & Color excess of the nebular lines (8 values linearly sampled)\\
   E(B-V)s factor      &  0.44     & Color excess ratio between continuum \& nebular  \\
   \hline
   \hline
   \label{tab: input_param}
   \end{tabular}
   \tablefoot{Where there is only one row per parameter, the same values were used to build the mock catalog and to fit. In the case of multiple rows, the first one was used to build the mock catalog, and the others were used to fit. This configuration resulted in $92~1600$ models per redshift.}
\end{table*}

\begin{figure}[!htbp]
\includegraphics[width=1\columnwidth]{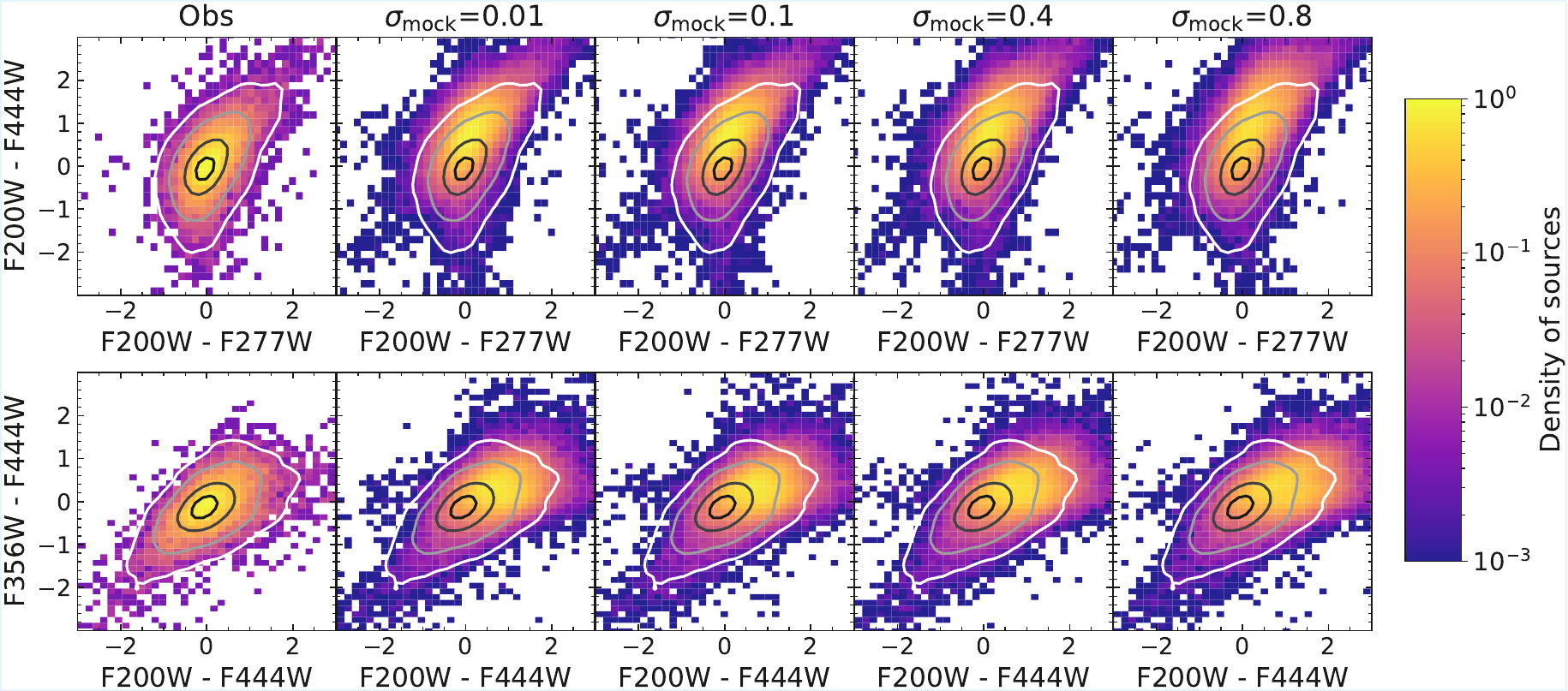}
\caption{Color-color diagram using JWST/NIRCam bands from the JADES catalog (left columns) and from the simulated catalogs (from second to last columns). These simulated catalogs are computed assuming, from left to right, $\sigma$ of 0.01, 0.1, 0.4, and 0.8. The two different rows show the color-color diagram obtained with two different band combinations chosen to take into account redshift and parameter coverage to ease the comparison. The color lines show the contours that enclose the 16$^{th}$, 50$^{th}$, 84$^{th}$, and 90$^{th}$ of the data in the observations. The implemented stochastic SFH can reproduce observed colors at $z>6$.}
\label{fig: color_dist}
\end{figure}

We built a mock catalog of galaxies at $6<z<12$ using the set of input parameters presented in Table~\ref{tab: input_param}, generating 10$^5$ simulated galaxies per $\sigma$. To compare with real observations, we computed the fluxes in the same set of filters as JADES \citep[see Sect.~\ref{sec: sample};][]{Eisenstein23b}, including 9 HST bands and 14 JWST/NIRCam bands. Then we perturbed the fluxes with a random Gaussian noise matching the sensitivity of JADES survey. 
In Fig.~\ref{fig: color_dist}, we show color distributions of the simulated catalog with added noise for $\sigma = 0.01,0.1,0.4,0.8$ compared to real observations (first column).  
The parameter range of the color distributions obtained with the simulated catalog is large enough to cover the observed colors for each burstiness level ($\sigma$), especially with $\sigma_{\rm mock} = 0.8$\footnote{Hereafter, we use the subscript ``mock'' to refer to mock catalogs generated with a given value of $\sigma$, and ``run'' to indicate the use of a fixed $\sigma$ value in the SED fitting process with \cigale.}. This first test ensures that the stochastic SFH, combined with typical SED modeling components, can reproduce a wide range of observed colors.

We used \cigale\ to fit the simulated SEDs. For this test, we adopted the input parameters listed in Table~\ref{tab: input_param}, that is the same set of parameters that we used to fit the observations. It is important to emphasize that we constructed one mock catalog for each of the seven tested $\sigma_{\rm mock}$ values. Each of these catalogs is then independently fitted seven times, each time fixing $\sigma_{\rm run}$ to a different value. This approach allows us to identify potential biases in the inferred SFHs when an inadequate model is used, as well as to assess the challenges in determining the best suited $\sigma_{\rm run}$. 

\begin{figure}[!htbp]
\centering\includegraphics[width=0.90\columnwidth]{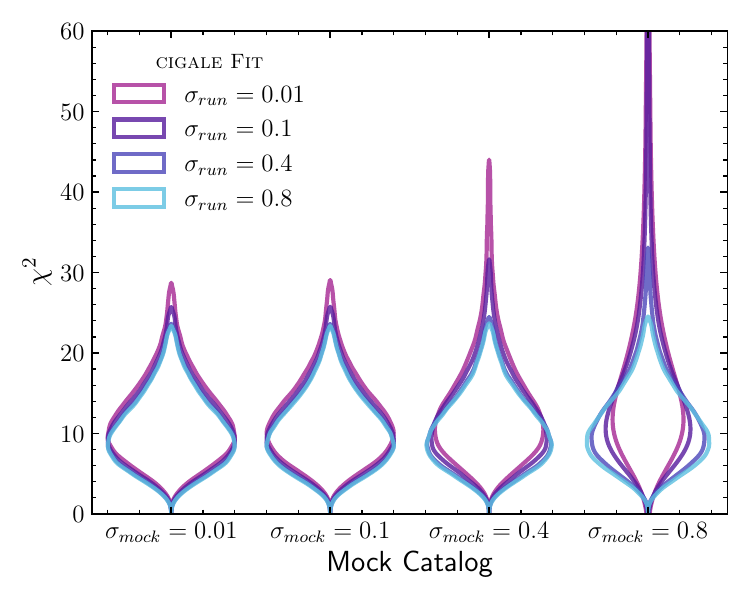}
\caption{$\chi^2$ obtained by fitting the mock catalogs with \cigale. The different fits for each simulated catalog are color-coded by $\sigma_{\rm run}$ used in the fit. A high level of burstiness ($\sigma_{\rm run}\sim0.8$) provides good fit quality, whatever the intrinsic level of burstiness of the fitted galaxy, while smoother SFHs struggle to fit bursty galaxy SEDs. 
}
\label{fig: violint_fit_mock}
\end{figure}

The $\chi^2$ distributions for these different configurations are shown in Fig.~\ref{fig: violint_fit_mock}. The distributions are narrow with a median value that depends on $\sigma_{\rm run}$. For all simulated catalogs, $\sigma_{\rm run}=0.8$, a high level of burstiness, shows the best results in terms of $\chi^2$ distribution. In contrast, using $\sigma_{\rm run}=0.01$ to fit galaxies' SED that were built with bursty SFH results in poorer fit quality as probed by $\chi^2$ distribution. This indicates that a model with a smooth evolution lacks the flexibility to accurately describe bursty SFHs.

\begin{figure}[!htbp]
\includegraphics[width=1\columnwidth]{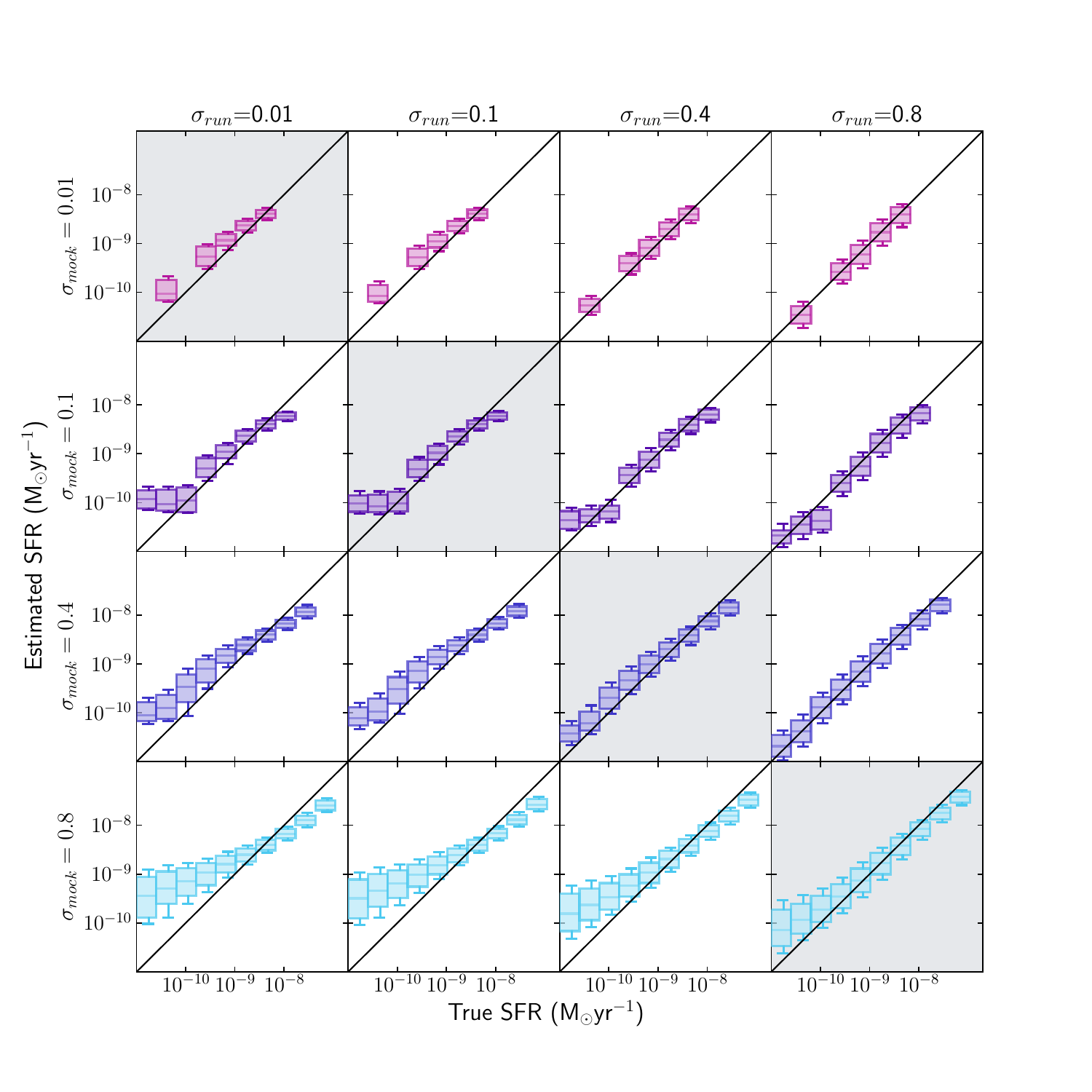}
\caption{Comparison between the mock SFR$_{\rm 10}$ (true SFR$_{\rm 10}$) and the SFR$_{\rm 10}$ recovered by \cigale\ (estimated SFR$_{\rm 10}$) for all mock catalogs (rows) fitted with the different values of $\sigma_{\rm run}$ (columns). The boxes span from the first quartile (Q1 = 25\%) to the third quartile (Q3 = 75\%), with a line indicating the median position, while the error bars extend from the 16$^{th}$ to the 84$^{th}$ percentile of the cumulative distribution. The gray background in the panels highlights when $\sigma_{\rm run}$ match $\sigma_{\rm mock}$. The SFH with the highest level of burstiness ($\sigma_{\rm run}=0.8$) provides the best estimate of SFR$_{\rm 10}$.}
\label{fig: sfr_fit_mock}
\end{figure}

We compare the SFR obtained through Bayesian-like analysis with \cigale, with the true values from the mock catalogs. In Fig.~\ref{fig: sfr_fit_mock}, for each combination of $\sigma_{\rm mock}$ and $\sigma_{\rm run}$, we show the recovered SFR as a function of the true ones. Note that the SFHs simulated by \cigale\ for the mock catalogs are normalized to 1\,M$_\odot$, making the resulting SFR values equivalent to the specific SFR (sSFR). The results suggest that the most bursty model ($\sigma_{\rm run}=0.8$) provides a better recovery of the SFR due to its greater flexibility in capturing both high and low values. This effect is particularly evident for the mock catalog built using $\sigma_{\rm mock}=0.8$, shown in the bottom row. When a smooth model (dominated by a $\tau$-delayed SFH, $\sigma_{\rm run}=0.01$) is used to fit galaxies with a stochastic SFH, the median SFR of galaxies with very low SFR ($<2.5\times10^{-11}$\,M$_{\odot}$yr$^{-1}$) is overestimated by $\sim1.3$\,dex. In contrast, when using the appropriate model, i.e., with ($\sigma_{\rm run}=0.8$), this difference is reduced by $\sim0.5$\,dex. At higher SFRs ($>4.3\times10^{-8}$\,M$_{\odot}$yr$^{-1}$), we observe an underestimation of $\sim0.5$\,dex when fitted with $\sigma_{\rm run}=0.01$, which decreases to $\sim0.25$\,dex when fitted with $\sigma_{\rm run}=0.8$. Furthermore, the bias in SFR estimation tends to decrease from $\sim0.7$\,dex to $\sim0.3$\,dex when a bursty SFH model is used.

These results show that the {\sc{StochasticSFH}} module implemented in \cigale\ in this work is able to reproduce a broad range of SFH and SFRs. In contrast, the classic parametric $\tau$-delayed model tends to overestimate the star formation rate and fails to capture extreme star-forming episodes, such as the bursts that galaxies at high redshift may undergo. It is important to note that this test was performed using the photometric filters covered in JADES (NIRCAM and HST). Including additional constraints, such as nebular emission lines such as H$\alpha$, although this is not accessible at our redshift range with JWST, is expected to further improve the accuracy of the results. 

\section{Observational data}\label{sec: sample}

We used the second data release\footnote{\url{https://archive.stsci.edu/hlsp/jades}.} (Program ID: 1180 and 1210, PIs: Eisenstein and  L\"utzgendorf, respectively) of the JWST Advanced Deep Extragalactic Survey \citep[JADES,][]{Rieke23,Bunker23,Eisenstein23,Hainline23}, which cover $\sim68$\,arcmin$^2$ of the \textit{Hubble} Ultra Deep Field, and presented in \cite{Eisenstein23b}. This catalog involves seven broad and five medium\footnote{F430M, F460M, and F480M images were acquired by JEMS but reprocess and release by JADES.} JWST/NIRCam bands from JADES (F090W, F115W, F150W, F200W, F277W, F335M, F356W, F410M, F430, F444W, F460M, and F480M), two medium NIRCam bands (F182M and F210M) from the JWST Extragalactic Medium-band Survey \citep[JEMS,][]{Williams23} and First Reionization Spectroscopically Complete Observations \citep[FRESCO,][]{Oesch23}, and nine HST bands (5 ACS + 4 WFC3) from the \textit{Hubble} Legacy Field \citep{Illingworth16}. The source detection was performed in an inverse variance-weighted stack image of the long wavelength bands (F277W, F335M, F356W, F410M, and F444W images), and the photometry was measured in Kron aperture and six different circular apertures. Photometric redshifts are computed with the code \texttt{EAZY} \citep{Brammer08}. The photometric redshifts have an average offset of 0.05 with a $\Delta z/(1+z)=0.024$ compared to a compilation of spectroscopic redshifts. For details on the data reduction process, photometry, and redshift estimates, we refer to \cite{Rieke23}, \cite{Eisenstein23b}, and \cite{Robertson24}. 

We used the photometric catalog based on KRON (PSF-Convolved)\footnote{The photometry was extracted in PSF-matched images convolved with NIRCAM F444W PSF.} apertures, restricted to NIRCam and HST/ACS bands, excluding ACS F850LP, as it is covered by the deeper NIRCam F090W data. HST/WFC3 photometry was not included due to its poor spatial resolution and low signal-to-noise ratio (S/N). The 5$\sigma$ flux depths for the broad NIRCam bands range from 30.7 to 30.2\,mag, while for the medium bands, they vary between 30.9 and 30.1\,AB\,mag. We selected sources within $6<z<12$ and imposed $\rm 23 \,mag < F277W < 30\,mag$ cut to remove contamination from stars and noisy photometry. When available, we adopted the spectroscopic redshifts from the third JADES data release\footnote{The third JADES data release for GOODS-S expanded the NIRSpec dataset through program PI:1210, providing over 2000 redshift measurements. However, this release does not include photometric data for GOODS-S.} presented in \cite{DEugenio25}. This selection results in a final sample of 6730 galaxies within $6<z<12$, 4369 objects at $6<z<7$, 1717 at $7<z<9$, and 644 at $z>9$. Spectroscopic redshifts account for only 1\% (73 sources) of the total sample.

\section{SED modeling with stochastic SFH}\label{sec: sed_modeling}

In this section, we perform SED fitting of the selected JADES subsample using the stochastic SFH described in Sect.~\ref{sec: stochastic_module}. 
Our goal is to test if a particular level of burstiness, characterized by $\sigma_{run}$, is more suitable to model the emission of $z>6$ galaxies and what its impact is on the estimated physical properties.

\subsection{Burstiness and fit quality}

We used \cigale\ to fit HST+JWST photometry and derive physical properties. The fitting setup follows the approach described in Sect.~\ref{sec: module_test}. We performed independent \cigale\ runs, each time varying the burstiness level ($\sigma_{\rm run} = 0.01, 0.1, 0.4, 0.6, 0.8, 1.0$), ranging from a classic $\tau$-delayed SFH with negligible stochasticity ($\sigma_{\rm run} \lesssim 0.1$) to a scenario dominated by strong bursts ($\sigma_{\rm run} = 1.0$). The input parameters used in the SED fitting are listed in Table~\ref{tab: input_param}.

\begin{figure}[!htbp]
\includegraphics[width=1\columnwidth]{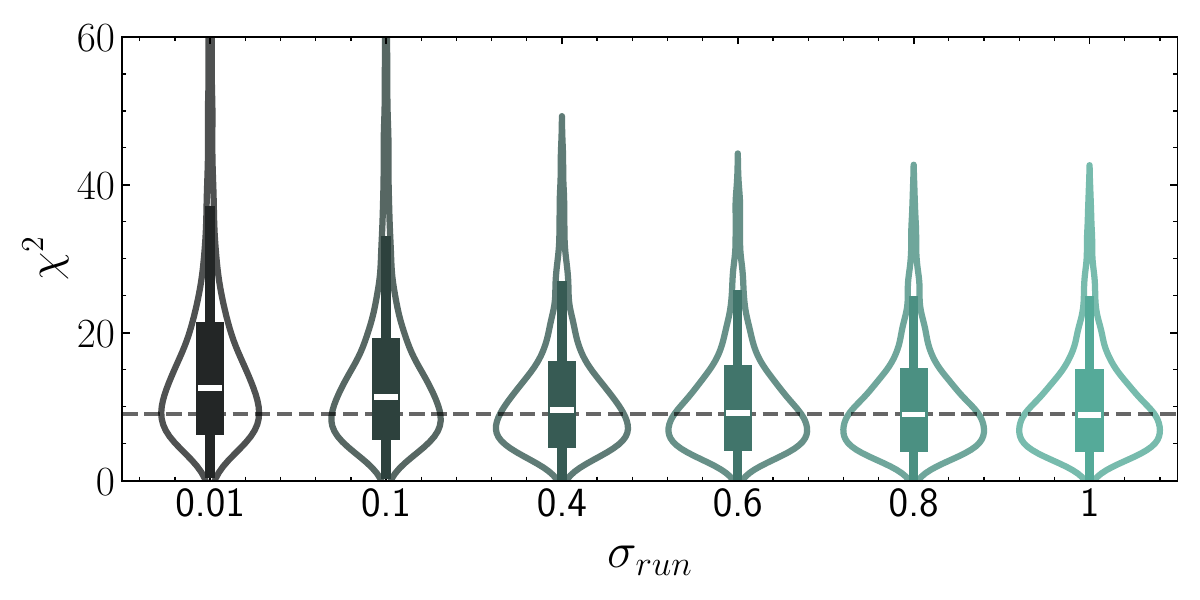}
\caption{$\chi^2$ distribution of the whole sample for each \cigale\ run. The $\sigma_{\rm run}$ used to model the SFH is indicated by the x-axis. The black dashed line shows the median $\chi^2$ for $\sigma_{run} = 0.8$. The embedded box plot of each distribution indicates the median, interquartile range (IQR), and whiskers. A high level of burstiness accounted for in the SFH is needed to fit early galaxies' SED.}
\label{fig: chi_2_distributions}
\end{figure}

In Fig.~\ref{fig: chi_2_distributions}, we compare the $\chi^2$ distribution obtained for the different $\sigma_{\rm run}$ values. The fit quality improves as the burstiness level increases. The distribution also becomes narrower at high $\sigma_{\rm run}$, reinforcing the preference for a stochastic SFH at high redshift. The median $\chi^2$ (white line in Fig.~\ref{fig: chi_2_distributions}) saturates at $\sim9$ when $\sigma_{\rm run}\gtrsim 0.8$ is reached. The improvement of fit with increasing stochasticity is also found in individual redshift bins. These results confirm the conclusion obtained from the analysis of the mock catalogs: a high level of burstiness accounted for in the SFH is needed to fit early galaxies' SED. This is also consistent with the conclusions of \cite{Faisst24}, who found that the best-fit parameter for galaxies at $z\sim7$ in the {\sc{Sphinx}}$^{20}$ simulation is $\sigma \sim 0.7$. 
The improvement of the $\chi^2$ distribution is the results of the wider variation in the recent SFH allowed by the highly bursty SFH. 
However, even if these $\chi^2$ distributions show that models with high burstiness level are needed to fit properly the observations, a lower $\chi^2$ obtained by a more complex model is not necessarily significant statistically as shown in \cite{Aufort20}.
Hence we do not rely on the $\chi^2$ distributions for our analysis in the rest of this work.

\subsection{Impact of the SFH burstiness on the main sequence}\label{sec: main_sequence}
\begin{figure}[!htbp]
\includegraphics[width=1\columnwidth]{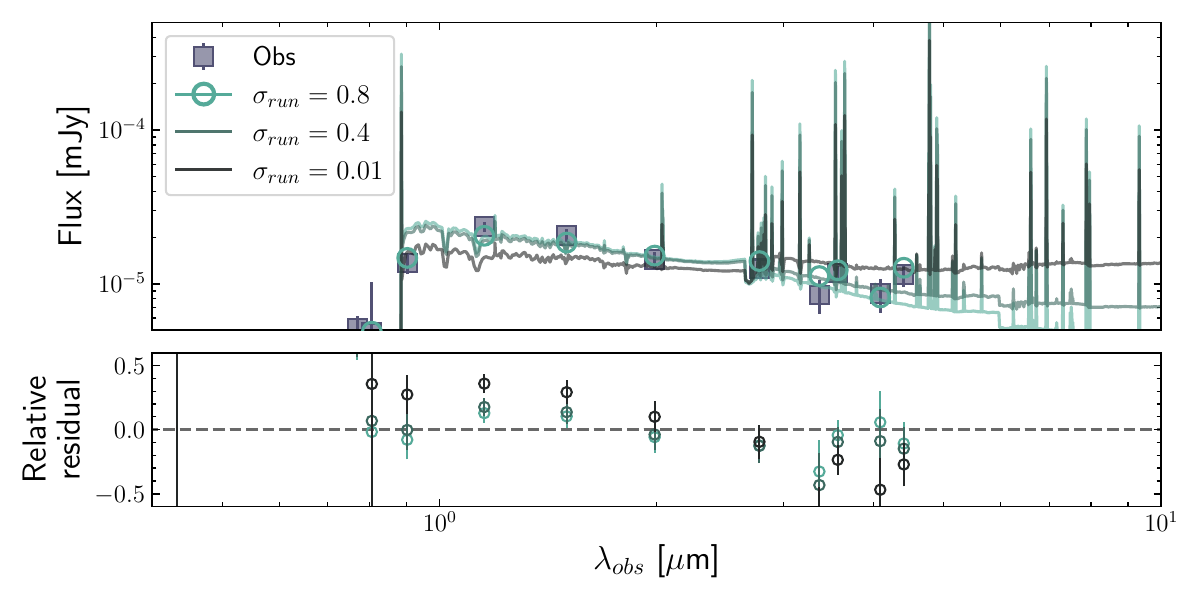}
\caption{Example SED of a galaxy at $z\sim6$ (top panel) and residuals (bottom panel). The dark purple squares are the fluxes from HST and JWST. The solid lines are the best-fit models obtained by \cigale. The open circles indicate the modeled fluxes for the different models (in the top panels only for $\sigma_{run}=0.8$).} 
\label{fig: SED}
\end{figure}

We now investigate the impact of the level of burstiness on the derivation of physical properties. As an example, we show in Fig.~\ref{fig: SED} the best-fit SED obtained with \cigale\ for a galaxy at $z\sim6.3$, for $\sigma_{\rm run} = 0.01,0.4$, and 0.8. The model using high $\sigma_{\rm run}$ provides a better fit of the observations, especially compared to the $\tau$-delayed model ($\sigma_{\rm run} = 0.01$). Although, the number of free parameters is the same for each model, the $\chi^2$ obtained decreases significantly as $\sigma_{\rm run}$ increases, dropping from 60.9 to 19.4 when comparing $\sigma_{\rm run} = 0.01$ and $\sigma_{\rm run} = 0.8$, respectively. The SED using $\tau$-delayed SFH ($\sigma_{\rm run} = 0.01$) underestimates fluxes at $<2$\,$\mu$m and yields a redder UV $\beta$ slope.

When considering the full galaxy sample, we find that the choice of SFH affects the derived properties (see Appendix~\ref{app: physical properties}). The stellar masses are higher by $\sim0.1$\,dex when low $\sigma_{\rm run}$ is assumed ($\lesssim$0.1). At higher redshift, the rest-frame NIR emission is less constrained (see Appendix \ref{app: stellar_mass} for further details) and the difference in stellar mass between the different levels of burstiness becomes negligible. The SFRs are higher by $\sim0.25$\,dex at $6<z<7$ to $\sim0.09$\,dex at $9<z<12$ for $\sigma_{\rm run}=0.8$. The UV $\beta$ slope is slightly bluer for $\sigma_{\rm run}=0.8$ compared to $\sigma_{\rm run}=0.01$ at $6<z<9$, while at $z>9$, the median values are nearly identical. However, the stochastic SFH ($\sigma_{\rm run}=0.8$) allows for a wider range of $\beta$ values in all redshift bins. 

\begin{figure*}[!htbp]
\centering
\includegraphics[width=1\textwidth]{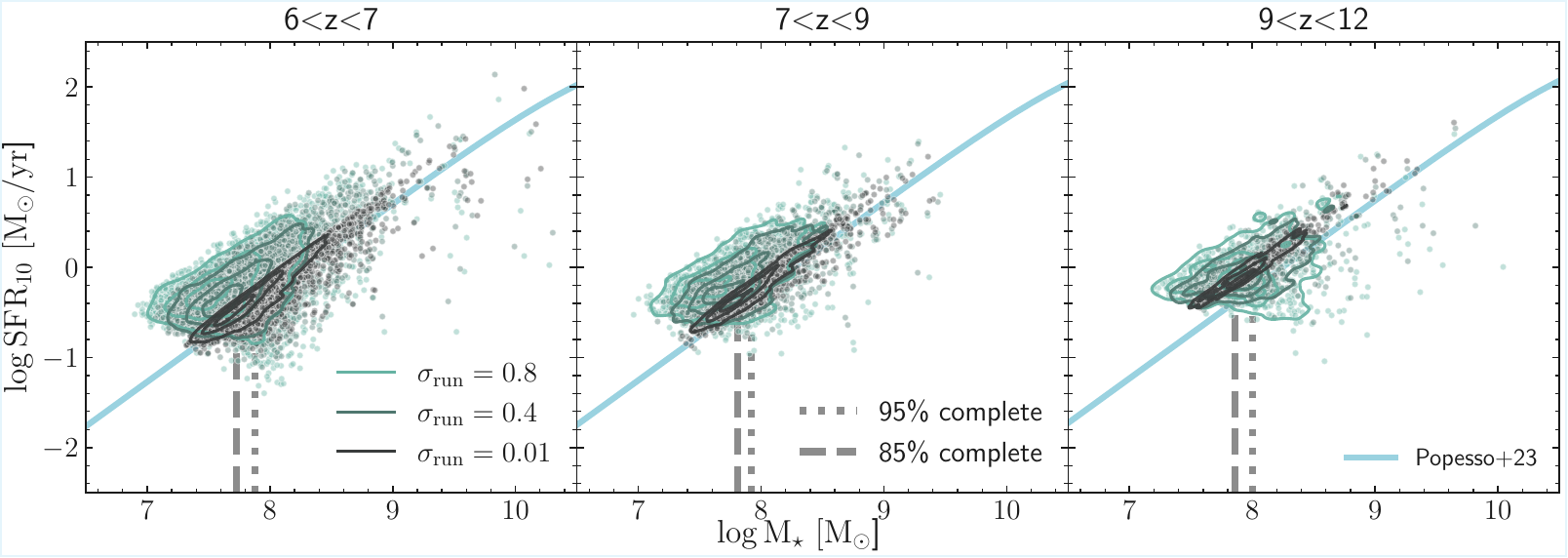}
\caption{SFR-M$_\star$ relation obtained from different burstiness levels in the SFH, color-coded by the assumed $\sigma_{\rm run}$. The dashed and dotted gray lines show when the classified sample (see Sect.~\ref{sec: galaxy_classification}) is 85\% and 95\% complete, respectively. The blue line shows the $\rm SFR-M_\star$ from \cite{Popesso23}, converted from a \cite{KroupaBoily02} to a \cite{Chabrier03} IMF. The scatter of the MS strongly depends on the burstiness encoded in the assumed SFH.}
\label{fig: main_squence_sfh_model}
\end{figure*}

To explore further the impact of the burstiness assumed by SFH models, we place the JADES galaxies of our sample in the $\text{SFR}-\text{M}_\star$ plane in Fig.~\ref{fig: main_squence_sfh_model}. In the three redshift bins, the $\tau$-delayed model ($\sigma_{\rm run}=0.01$) results in a strong and tight correlation between the SFR and the stellar mass, even at high redshifts ($z>9$). This is due to the fact that this model offers a limited range of SFR achievable \citep[e.g.,][]{Ciesla17}. The dispersion around the MS increases with increasing $\sigma_{\rm run}$. In Table~\ref{tab: spearman_coefficients_main_sequence}, we compute the Spearman coefficients for the full sample and for different SFH models. With $\sigma_{\rm run}=0.01$, the correlation is very strong $>0.8$ and reaches the high value of 0.9 in the $9<z<12$ bin. 
Smooth SFH models impose a strong prior on sSFR, limiting the range of possible SFRs and producing a well-defined main sequence \citep{Buat14, Carnall19}. In Sect.~\ref{sec: module_test}, we observed similar trends when fitting the mock catalogs. Models with low burstiness levels lack flexibility, often leading to overestimated or underestimated SFRs in bursty galaxies. 
At higher burstiness level, the Spearman coefficient decreases with increasing $\sigma_{\rm run}$ and increasing redshift.
For $\sigma_{\rm run}=0.8$, these coefficients are indicating a moderate to weak correlation between SFR and M$_\star$ at $z\geq7$ while the relation remains strong at $z<7$.
These results are consistent with \cite{Ciesla24}, who found similar coefficient ranges and trends with redshift, although using a nonparametric SFH (flat prior, linearly sampled time bins). 
Our results indicate that the MS scatter is strongly dependent on the choice of SFH model, which has already been discussed in the literature \citep[e.g.,][]{Buat14, Ciesla17, Leja22}, and the level of burstiness modeled in it.

\begin{table}[!htbp]
    \centering
    \caption{Spearman correlation coefficients for the $\text{SFR}_{10}-\text{M}_*$ plane with the different SFH models ($\sigma_{\rm run}$) in the three redshift bins.}
       \begin{tabular}{ccccc}
    \hline 
    \hline 
    \multirow{2}{*}{Redshift bin} & \multicolumn{4}{c}{Spearman Coefficient} \\
    \cline{2-5}
     & $\rm \sigma_{run}=0.01$ & $\rm \sigma_{run}=0.1$ & $\rm \sigma_{run}=0.4$ & $\rm \sigma_{run}=0.8$ \\
    \hline 
    \hline
    {$6<z<7$} & 0.80 & 0.74 & 0.56 & 0.46 \\
    {$7<z<9$} & 0.86 & 0.81 & 0.57 & 0.44 \\
    {$9<z<12$} & 0.93 & 0.89 & 0.54 & 0.21 \\

    \hline  
    \hline 
    \end{tabular}
    \label{tab: spearman_coefficients_main_sequence}
\end{table}

\section{SFH classification}\label{sec: galaxy_classification}

We investigate whether the galaxies in our sample exhibit stochastic star formation, based on the result of the SED fitting assuming different burstiness. As explained in Sect.~\ref{sec: intro}, there is increasing evidence from observations that galaxies at high redshift experience random periods of star formation \citep[e.g.,][]{Dressler24,Endsley24b,Ciesla24,Langeroodi24a,Cole25}. We first classify galaxies according to their preferred level of burstiness obtained from the SED fitting. We then select galaxies to build a mass-complete sample. We finally study the evolution of the fraction of bursty galaxies and their physical properties.

\subsection{Galaxy classification with Bayes factor}\label{sec: bayes_factor}

We compare the quality of fit obtained from the \cigale\ runs using the different levels of burstiness.
We thus use the Bayes factor (BF) defined as:  
\begin{equation}
    BF_{\frac{\sigma_{\rm run}=0.8}{\sigma_{\rm run}=X}} = \frac{p(x_{obs}|(\sigma_{\rm run}=0.8))}{p(x_{obs}|(\sigma_{\rm run}=\rm X))},
\end{equation}
where $p(x_{obs}|\text{m})$ is the likelihood integrated over the prior distribution for a given model $m$ and observations $x_{obs}$. We choose the model with  $\sigma_{\rm run}=0.8$, as our reference model, to be able to account for high burstiness, as smooth SFH models can not reproduce bursty galaxies, according to $\chi^2$ distribution (Fig.~\ref{fig: chi_2_distributions}). The BF is then computed for each other model against this reference. We also implemented some model quality control: for each model, we first exclude observations for which the fit yields an effective sample size ($n_{\mathrm{eff}}$) below 100. The effective sample size estimates how many independent samples effectively contribute to the posterior distribution approximation; a low $n_{\mathrm{eff}}$ indicates that the sampling process is inefficient and dominated by low-weight samples. In such cases, the likelihood is effectively informed by too few independent models and the resulting Monte Carlo approximation of the integral is poor.

We used $\sigma_{\rm run}=0.8$ as reference and define it as ``bursty SFH." To be conservative, we define as ``smooth" the run made with $\sigma_{\rm run}=0.01$. We now divide galaxies in our sample in two groups: bursty galaxies ($\rm BF_{\sigma_{\rm run}=0.8/\sigma_{\rm run}=0.01}=BF_{Bursty/Smooth} > 3$) and smooth galaxies ($\rm BF_{Bursty/Smooth} < 3$), following the Jeﬀreys' scale \citep[see, eg.,][]{Robert07, Aufort20}. 
For galaxies classified as smooth ($\rm BF_{Bursty/Smooth}<3$), we assign the physical properties obtained from the run with $\sigma_{\rm run}=0.01$. This classification results in 30\% (1970) of galaxies being identified as bursty and 70\% (4600) as smooth in the whole sample. We test the impact of this classification by changing the reference burstiness model from $\sigma_{\rm run}=0.8$  to $\sigma_{\rm run} = 0.6$, finding that the respective percentages vary by only $\sim 1\%$. In Appendix~\ref{app: clasification_test}, we test the classification method and examine the impact of the S/N and the choice of the BF threshold ($\rm BF_{Bursty/Smooth} > 3$).

\subsection{Mass-complete sample and physical properties}\label{sec: mass_complet}

For the rest of the analysis, we need to determine the mass-completeness limit of our sample following the approach proposed by \citet{Pozzetti10} \citep[see also][]{Florez20,Mountrichas22,Ciesla24}. We estimated the limiting stellar mass ($\rm log M_{\star,lim}$) of each galaxy, defined as the stellar mass a galaxy would have if its apparent magnitude (m$_{AB}$) were equal to the limiting magnitude of the survey (m$_{AB,lim}$), $\rm log M_{\star,lim} = log M_{\star} + 0.4 \times (m_{AB}-m_{AB,lim})$. This gives us the distribution of limiting stellar masses for our sample in each redshift bin. To define a representative mass-completeness limit, we selected the faintest 20\% of galaxies in each redshift bin and computed the 80$^{th}$ and 95$^{th}$ percentiles of their $\rm log M_{\star,\mathrm{lim}}$ distribution. In this work, we used the NIRCam/F277W filter and the selection cut we imposed in Sect.~\ref{sec: sample} as the limiting magnitude, 30\,mag, which is brighter than the sensitivity of the survey, $\sim30.7$\,mag \citep{Eisenstein23b}. 
The limiting stellar mass could be affected by the choice of SFH. However, at $z<9$, the difference in stellar mass between models is below 0.06\,dex, and $<0.25$\,dex in the highest redshift bin. Our sample is 85\% complete above $\rm log(M_{\star,\mathrm{lim}}) =$ 7.73, 7.81, and 7.86 at $6<z<7$, $7<z<9$, and $9<z<12$, respectively, and 95\% complete above $\rm log(M_{\star,\mathrm{lim}}) =$ 7.84, 7.92, and 8.0 in the same redshift bins (see Fig.~\ref{fig: main_squence_sfh_model}). From now on, we only study galaxies with stellar masses above the limits, ensuring 95\% mass-completeness.
In Table~\ref{tab: prop_galaxies_complete_sample}, we provide the median properties of the mass-complete sample.

\begin{table*}
    \centering
    \caption{Properties derived from SED modeling of the mass-complete sample assuming $\sigma_{\rm run} = 0.01, 0.4$, and 0.8.}
    \renewcommand{\arraystretch}{1.2}
    \begin{tabular}{c c c c c c c c}
        \hline
        \hline
         & $z$ & $\sigma_{\rm run}$ & $\chi^2$ & $\log_{10}$M$_\star$ ($\rm M_\odot$) & $\log_{10}$ SFR$_{10}$ ($\rm M_\odot$/yr) & $\beta$ &  ${\rm M_{UV}}$ (mag) \\
        \hline
        \hline 
        \multirow{3}{*}{Sample} & \multirow{3}{*}{$6<z<7$} 
        & 0.01 & $12.9_{-6.3}^{+11.2}$ & $8.10_{-0.14}^{+0.30}$ & $-0.14_{-0.21}^{+0.31}$ & $-2.32_{-0.07}^{+0.12}$ & $-18.30_{-0.74}^{+0.68}$ \\
        & & 0.4 & $9.8_{-4.7}^{+8.7}$ & $8.10_{-0.14}^{+0.35}$ & $0.04_{-0.39}^{+0.36}$ & $-2.31_{-0.13}^{+0.14}$ & $-18.44_{-0.88}^{+0.89}$ \\
        & & 0.8 & $9.2_{-4.5}^{+8.0}$ & $8.10_{-0.14}^{+0.37}$ & $0.10_{-0.51}^{+0.40}$ & $-2.30_{-0.15}^{+0.17}$ & $-18.51_{-0.95}^{+1.04}$ \\
        \hline
        \multirow{3}{*}{Sample} & \multirow{3}{*}{$7<z<9$} 
        & 0.01 & $11.9_{-5.5}^{+10.4}$ & $8.09_{-0.14}^{+0.30}$ & $-0.08_{-0.16}^{+0.29}$ & $-2.30_{-0.09}^{+0.10}$ & $-18.30_{-0.77}^{+0.48}$ \\
        & & 0.4 & $9.4_{-4.4}^{+7.6}$ & $8.09_{-0.14}^{+0.35}$ & $0.07_{-0.29}^{+0.32}$ & $-2.30_{-0.12}^{+0.12}$ & $-18.48_{-0.77}^{+0.67}$ \\
        & & 0.8 & $8.6_{-4.1}^{+7.1}$ & $8.12_{-0.17}^{+0.37}$ & $0.12_{-0.42}^{+0.36}$ & $-2.28_{-0.15}^{+0.14}$ & $-18.52_{-0.84}^{+0.80}$ \\
        \hline
        \multirow{3}{*}{Sample} & \multirow{3}{*}{$9<z<12$} 
        & 0.01 & $10.21_{-4.4}^{+6.7}$ & $8.20_{-0.16}^{+0.28}$ & $0.17_{-0.16}^{+0.26}$ & $-2.29_{-0.10}^{+0.10}$ & $-18.77_{-0.77}^{+0.45}$ \\
        & & 0.4 & $9.1_{-4.2}^{+6.0}$ & $8.20_{-0.14}^{+0.34}$ & $0.09_{-0.26}^{+0.35}$ & $-2.23_{-0.13}^{+0.12}$ & $-18.63_{-0.85}^{+0.61}$ \\
        & & 0.8 & $8.9_{-4.1}^{+6.2}$ & $8.22_{-0.17}^{+0.35}$ & $-0.06_{-0.31}^{+0.43}$ & $-2.20_{-0.15}^{+0.17}$ & $-18.49_{-0.97}^{+0.62}$ \\
        \hline
        \hline
    \end{tabular}
    \label{tab: prop_galaxies_complete_sample}
\end{table*}

\section{Evolution of galaxies' burstiness}\label{sec: evolution_gal_burstiness}

\subsection{From the SFH classification}

\begin{figure*}[!t]
 \centering\includegraphics[width=0.48\textwidth]{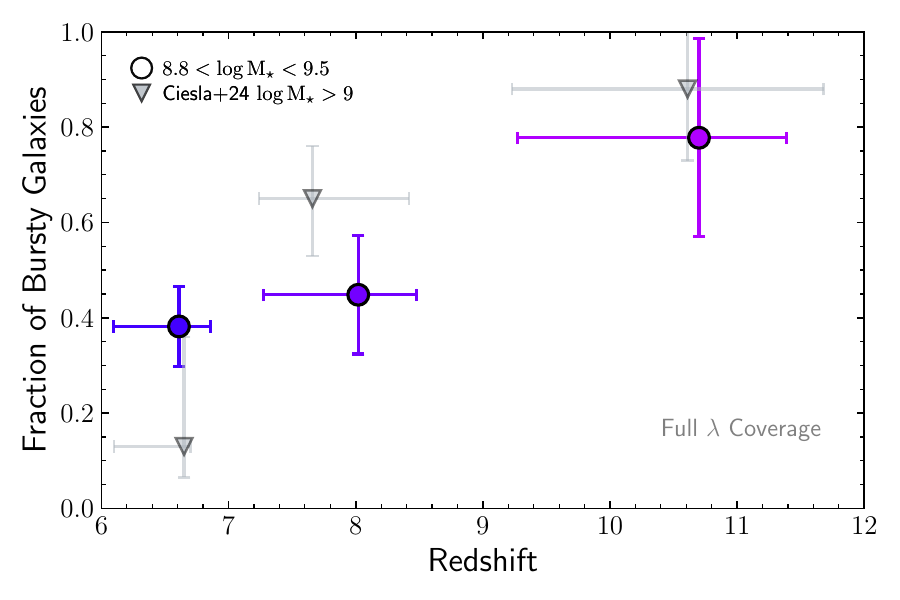} \includegraphics[width=0.48\textwidth]{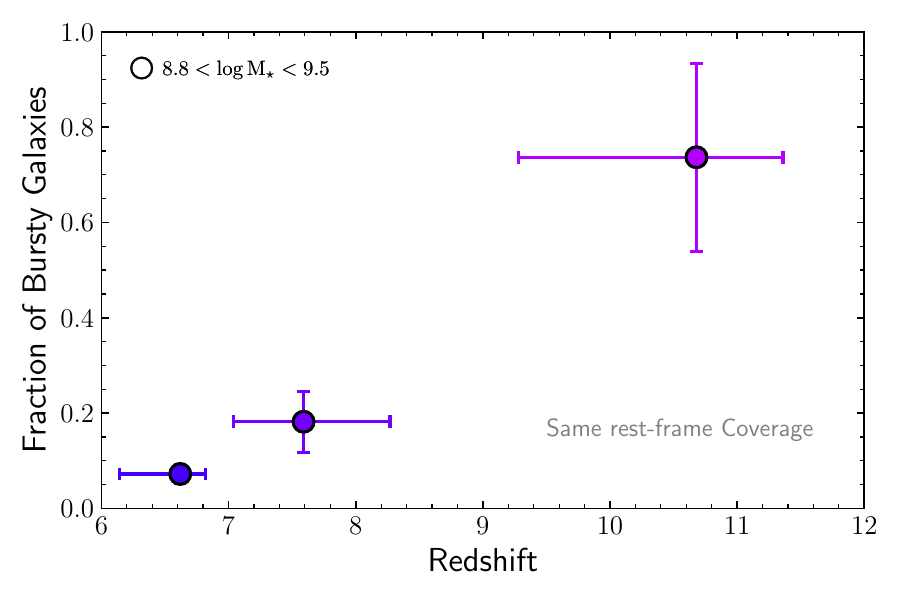}
\caption{Fraction of high-mass ($8.8 < \log M_\star/M_\odot < 9.5$) galaxies classified as bursty as a function of redshift using the Bayes factor (see Sect.~\ref{sec: bayes_factor}).
The left panel shows the results when full spectral information at each redshift is used, while the right panel shows the results assuming the same rest-frame wavelength coverage as a galaxy at $z=11$. Uncertainties on the fractions are derived from Poisson distribution, while uncertainties on the redshifts correspond to the 16$^{th}$–84$^{th}$ percentile range of the redshift distribution within each stellar mass bin. The gray downside triangles show the fraction of stochastic massive galaxies ($\rm \log M_\star>$9) found in \cite{Ciesla24}, obtained from the variations of the recent SFR using nonparametric SFH. There is a trend showing an increasing fraction of bursty galaxies with increasing redshift that is not due to the difference in wavelength coverage.}
\label{fig: f_bursty_galaxies}
\end{figure*}

The fraction of galaxies in the two groups described in Sect.~\ref{sec: bayes_factor}, bursty and smooth, evolves with redshift. 
In the left panel of Fig.~\ref{fig: f_bursty_galaxies}, we show the fraction of high-mass galaxies (8.8 $< \log M_\star/M_\odot <$ 9.5) with a bursty SFH, according to the BF, as a function of redshift in the three redshift bins considered in this work. We choose this number to be 1\,dex above the mass-completeness limit, to ensure a 100\% complete sample. There is a tentative trend indicating an increasing fraction of high-mass galaxies with a bursty SFH with redshift. The trend is not clear between $z\sim6.5$ and $z\sim8$ with consistent median values, although the dispersion is smaller in the first redshift bin. At $z>8$, the fraction of galaxies classified as bursty strongly increases with redshift. The overall trend is consistent with the results of \cite{Ciesla24} obtained from a different method (distribution of the variations of the recent SFH using nonparametric models) on the same sample. 
To understand if the different rest-frame wavelength coverages due to redshift probed by the HST+JWST observations could bias our results in artificially creating this trend, we performed a second \cigale\ run removing longer wavelength filters in galaxies with $z<11$ in order to match the rest-frame wavelength coverage of galaxies at $z=11$, although this is not ideal as the wavelength coverage of the filters is not homogeneous. In the right panel of Fig.~\ref{fig: f_bursty_galaxies}, we show the resulting fractions of bursty galaxies obtained with this limited wavelength coverage. The trend is stronger, indicating that the difference in rest-frame coverage is not the origin of the decrease of the fraction of bursty galaxies with decreasing redshift.

For galaxies with lower stellar masses ($\log M_\star/M_\odot < 8.8$), we obtain a fraction of bursty galaxies lower than 20\% at all redshifts. This could be surprising as it is expected that galaxies with lower stellar mass have more bursty SFH \citep[e.g.,][]{Langeroodi24c}. Low-mass galaxies are fainter with an average $M_{\mathrm{UV}}$ of $-18.3_{-0.6}^{+0.9}$\,mag in this mass bin compared to $-20.1_{-0.8}^{+0.9}$\,mag in the high-mass bin. Therefore, the uncertainties in the flux densities are larger: $15\%$ on average in the JWST/F277W filter for the low-mass bin compared to 7\% for the high-mass bin. To test whether this could be the origin of this low fraction of bursty sources, we use the best-fit models obtained with $\sigma_{\mathrm{run}} = 0.8$ as realistic simulated observations to evaluate how the classification is impacted by the S/N in Appendix~\ref{app: clasification_test_snr}. We find that the fraction of bursty galaxies correctly identified drops from $\sim70\%$ at S/N~$\sim15$ to $\sim32\%$ at S/N~$\sim7$ (typical for the low-mass bin), implying that a large fraction of low-mass bursty systems can be misclassified by the BF analysis due to their low S/N. We note that this is partly driven by our conservative classification threshold ($\mathrm{BF}_{\mathrm{Bursty/Smooth}} < 3$), which is set to obtain a pure sample of bursty galaxies at $\sim$95\% (see Appendix~\ref{app: clasification_test_snr}).
Hence, it is difficult to distinguish between models for faint sources from the BF analysis, as bursty and smooth SFH result in equivalently good $\chi^2$, confirming the importance of S/N in our analysis.

\subsection{The $\rm \log(SFR_{10}/SFR_{100})$ ratio}

\begin{figure*}[!htbp]
\centering\includegraphics[width=1\textwidth]{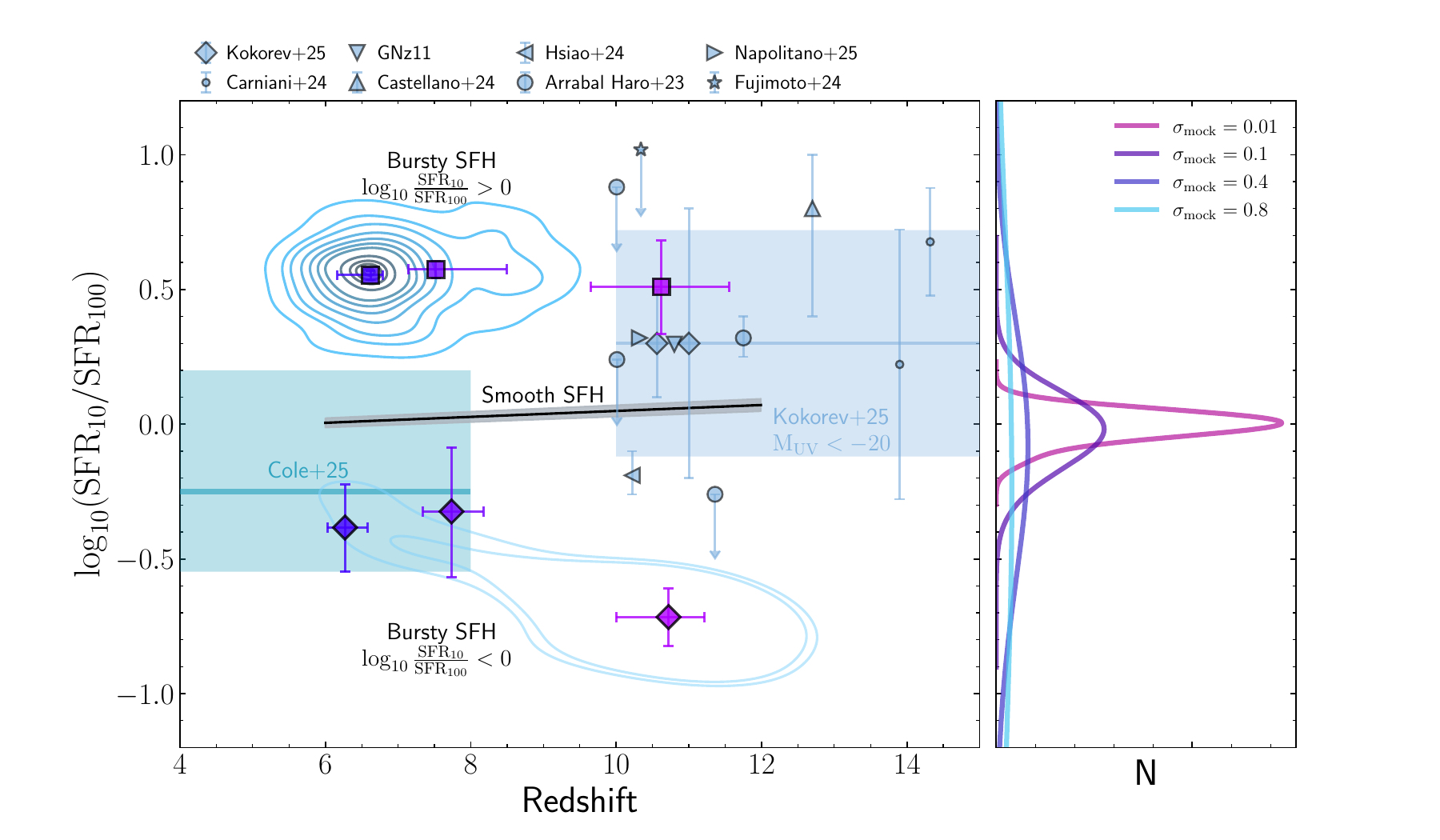}
\caption{Evolution of the $\rm \log_{10}(SFR_{10}/SFR_{100})$ ratio, used as an indicator of burstiness, as a function of redshift. Blue contours indicate the distribution of the galaxies of our sample classified as having a bursty SFH. The big square and diamonds indicate the median value for each of the three redshift bins considered in this work for bursty galaxies with enhanced SF activity and decreasing activity, respectively. The black solid line is a linear fit to the ratio obtained for galaxies classified as having a smooth SFH. Measurements from the literature are also shown for comparison \citep{Cole25, Kokorev25, Carniani24}. The $\rm SFR_{10}/SFR_{100}$ ratio from \cite{ArrabalHaro23, Castellano24, Hsiao24, Bunker23b, Alvarez25, Fujimoto24, Napolitano25} were taken from \cite{Kokorev25}. The light blue shaded regions represent the $\rm \log_{10}(SFR_{10}/SFR_{100})$ distribution within 1-$\sigma$ for bright galaxies \citep[M$_{\rm UV}<-20$;][]{Kokorev25} and the 16$^{th}$–84$^{th}$ percentiles from \cite{Cole25}, with the median indicated by the solid line. The right panel shows the distribution of $\rm \log_{10}(SFR_{10}/SFR_{100})$ ratios obtained from the simulated mock catalogs with different burstiness levels presented in Sect.~\ref{sec: module_test}. The ratio is constant as a function of redshift. Only high ratios of $\gtrsim0.2$ cannot be reproduced by a smooth SFH. } 
\label{fig: log10_sfr10_sfr1000_z}
\end{figure*}

Another way to probe the burstiness of galaxies' star formation activity is to look at the ratio between the most recent SFR, for instance averaged over the last 10\,Myr, and the SFR averaged over a longer timescale such as 100\,Myr \citep[e.g.,][]{Endsley24b,Langeroodi24c,Kokorev25}. The most recent SFR is traced by emission lines, while the longer timescales are probed by continuum. However, we only have access to spectroscopy for 1\% of our sources. To understand if we recover $\rm \log_{10}(SFR_{10}/SFR_{100})$ that are consistent with studies from the literature using spectroscopy, we show this ratio as a function of redshift in Fig.~\ref{fig: log10_sfr10_sfr1000_z}. We separate galaxies classified as having a smooth SFH from the bursty galaxies, and we highlight the median value of the ratio for bursty galaxies with enhanced SFR and bursty galaxies that underwent a decrease of their star formation activity. Overall, our estimates are in agreement with values from the literature obtained as statistical samples \citep{Cole25} or from individual sources \citep{ArrabalHaro23,Carniani24,Castellano24,Hsiao24,Fujimoto24,Napolitano25,Kokorev25}. Galaxies of our samples tagged with smooth SFH show a slight decrease of the $\rm \log_{10}(SFR_{10}/SFR_{100})$ ratio with decreasing redshift. This trend is consistent with those of \cite{Kokorev25} at $z>14$ and \cite{Cole25} at $z<8$, bridging these two relations. Galaxies classified as bursty with an enhanced SF activity ($\rm \log_{10}(SFR_{10}/SFR_{100})>0$) have a typical ratio of $\sim$0.5 that is consistent with the individual sources at $z>10$ as well as the relation from \cite{Kokorev25}. We do not see any significant decrease of the median $\rm \log_{10}(SFR_{10}/SFR_{100})$ ratio with decreasing redshift for these galaxies.
Interestingly, $\sim$12\% of the galaxies in our sample that are classified as bursty show a negative $\rm \log_{10}(SFR_{10}/SFR_{100})$ ratio, indicating that they recently underwent a rapid decrease of their SF activity. These galaxies are expected and consistent with individual sources that have been identified in the literature as lull or mini-quenched galaxies \citep{Dome24,Looser24a,Endsley24b,Looser25,CoveloPaz25}.

In Fig.~\ref{fig: log10_sfr10_sfr1000_z}, we place our results for the JADES sample along with measurements from the literature using different surveys and different types of data (photometric and/or spectroscopic). In these studies, the $\rm \log_{10}(SFR_{10}/SFR_{100})$ parameter is used as an indicator of burstiness and galaxies with $\rm \log_{10}(SFR_{10}/SFR_{100})>0$ are classified as bursty \citep{Endsley24b,Kokorev25}.
Gathering a sample of 12 galaxies at $z>10$ with spectroscopic information, \cite{Kokorev25} found a mean $\rm \log_{10}(SFR_{10}/SFR_{100})$ of 0.30$\pm$0.02\,dex and a scatter of 0.42$\pm$0.02\,dex around this value.
At lower redshift, $4<z<6$, \cite{Cole25} obtained a mean value of $\sim-0.21$  with a large scatter of 0.46. 
The combination of these results would point toward a decrease of the burstiness of galaxies with decreasing redshift. 
We use the simulated mock sample of galaxies presented in Sect.~\ref{sec: module_test} with different $\sigma_{mock}$ values to derive the distributions of $\rm \log_{10}(SFR_{10}/SFR_{100})$ and show them in the right panel of Fig.~\ref{fig: log10_sfr10_sfr1000_z}.
A pure $\tau$-delayed SFH only yields a narrow distribution of $\rm \log_{10}(SFR_{10}/SFR_{100})$ around 0.
However, an SFH with a low level of burstiness $\sigma_{mock}=0.1$ (see Fig.~\ref{fig: SSFH}) can produce a distribution with SFH models reaching $\rm \log_{10}(SFR_{10}/SFR_{100})$ values of $\pm$0.3. 
Values exceeding these limits can only be reproduced by bursty SFH models.
Hence, a positive value of $\rm \log_{10}(SFR_{10}/SFR_{100})$ does not necessarily imply a bursty SFH if the value is close to 0 ($\pm$0.2-0.3), but a higher ratio would definitely rule out a smooth SFH. 

To investigate a possible evolution of the burstiness level of the galaxies in our sample, we compute the fraction of massive galaxies (8.8 $<\log M_\star/M_\odot <$ 9.5), for consistency with the previous section, that have a $\rm \log_{10}(SFR_{10}/SFR_{100})$ larger than 0.2. 
At $z>9$, no massive sources have a ratio larger than 0.2. 
However, we obtain a fraction of 0.28$\pm$0.06 and 0.38$\pm$0.11, at $6<z<7$ and $7<z<9$, respectively.
Although the Poissonian errors are large, especially at $7<z<9$, these numbers point toward an increasing fraction of bursty galaxies with increasing redshift, confirming the results of the previous section.

\section{Evolution of $\sigma_{UV}$}\label{sec: sigma_uv_redshift}

\begin{figure}[!htbp]
\centering\includegraphics[width=1\columnwidth]{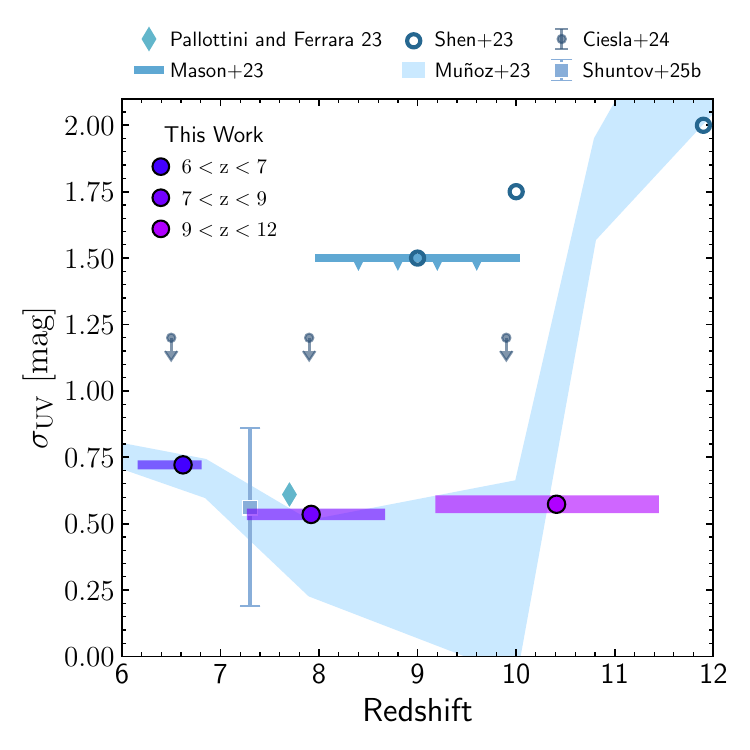}
\caption{Evolution of $\sigma_{UV}$ with the redshift for our sample in JADES at $\rm 6<z<12$. The uncertainties on $\sigma_{UV}$ are estimated from bootstrapping. The error bar on redshifts indicates the 16$^{th}$ to 84$^{th}$ percentiles of the distributions. 
The blue shades and markers show estimations from the literature, which are based on semi-analytical models and simulations \citep{PallottiniFerrara23, Shen23, Munoz23, Mason23} or observations \citep{Ciesla24, Shuntov25}. Our estimate $\sigma_{\rm UV}$ is consistent with the literature and too low to explain the mild evolution of the UVLF at $z>10$ only from bursty SFH, as estimated by \cite{Shen23} (empty circles) using empirical models.}
\label{fig: sigma_uv}
\end{figure}

In the literature, $\sigma_{UV}$ is defined as the dispersion in mag around a median $M_{\rm UV}$ obtained from a $M_{\rm UV}-M_{\rm halo}$ relation and has been used to quantify the stochasticity of star formation \citep{Shen23, Shuntov25}. To compute $\sigma_{\rm UV}$, we used the complete sample and estimated the median $M_{\rm UV}$ in bins of stellar masses to remove the trend between $M_{\rm UV}$ and $M_{\star}$. We normalized $M_{\rm UV}$ using the corresponding median value of their stellar mass bin and computed the dispersion of the corrected $M_{\rm UV}$ distribution.  
To account for uncertainties, we perform Monte Carlo realizations of both $M_{\star}$ and $M_{\rm UV}$ by drawing from Gaussian using the uncertainties from the SED fitting as the width of the distribution. In each realization, we apply the binning, normalization, and compute the scatter of the corrected $M_{\rm UV}$ distribution. The final value of $\sigma_{\rm UV}$ and its uncertainties are estimated from the percentiles of the resulting distribution.
We obtain $\sigma_{\rm UV}$ of 0.72$\pm$0.02, 0.54$\pm$0.02, and 0.57$\pm$0.03\,mag, at $6<z<7$, $7<z<9$, and $9<z<12$, respectively. We compare these values with results from the literature in Fig.~\ref{fig: sigma_uv}. 

Our measurements of $\sigma_{\rm UV}$ show relatively good agreement with previous results at $z\sim6-12$. \citet{Ciesla24} performed a similar analysis using SED modeling with a nonparametric SFH, and found $\sim 0.6$\,mag higher values when derived from the $\rm M_{UV}$ distribution. However, in their measurement, they took into account all galaxies with $\log M_\star/M_\odot \geq 9$, hence considering one mass bin. For this reason, their values can be considered as upper limits. 
A different observational approach is used in \cite{Shuntov25}, where a spectroscopic sample was used to constrain the M$_{\rm UV}$-M$_h$, the relation between star formation efficiency (SFE) and M$_h$, stochasticity and galaxy bias. Our results are consistent within their uncertainties. 
\cite{Munoz23} followed similar semi-empirical models to fit the UVLF and derived $\sigma_{\rm UV}$. They found low values $<0.8$\,mag, which match our results, followed by a sharp increase at $z\sim 10$ that can still be compatible with our results, although our high-$z$ bin value is low. Interestingly, at lower redshifts, our measurements follow well their predicted redshift dependence of $\sigma_{\rm UV}$. \cite{Mason23} used empirical models to study the $\rm M_{UV}$-M$_h$ relation and the impact of young bright galaxies in the UVLF, which are the most likely to dominate the observations, and showed an upscatter up to 1.5\,mag above the median relation. With our observations at the same redshift range, we observe a lower level of stochasticity in our $\rm \sigma_{UV}$ measurements. In the {\sc serra} simulations, \cite{PallottiniFerrara23} found $\sigma_{\rm UV}=0.61$\,mag at $z\sim7.7$, which is consistent with our results. Finally, \cite{Shen23} showed that to explain the observed UVLF at $z>10$, values as high as $\sigma_{\rm UV}>1.5$\,mag would be required. Our level of stochasticity, as probed by $\sigma_{\rm UV}$ is not high enough to explain the observed UVLF. 

Our results do not indicate an increase in $\sigma_{UV}$ that some works find necessary to describe the high-$z$ UVLF and thus reinforce the idea of a solution combining several effects in addition to bursty SFH. A high SFE has been found to play a role in the mild evolution of the UVLF at $z>10$ as discussed and found by \cite{Gelli24} and \cite{Shuntov25}. 
Here, we used our results, particularly the SFR obtained from specific modeling of the SFH to have a first order estimate of the instantaneous SFE ($\epsilon$\footnote{The instantaneous SFE is different from the integrated SFE ($\epsilon_{\star}=M_{\star} M_{\mathrm{h}}^{-1} f_{\mathrm{b}}^{-1}$).}, note that is different from the integrated star formation efficiency $\epsilon_{\star}$) by assuming the stellar-to-halo mass relation (SHMR) from \cite{Shuntov25a}. They inferred it using nonparametric abundance matching of their stellar mass function (SMF) and the \cite{Watson13} halo mass function (HMF). This allowed us to estimate the halo mass of our sample, then compute the halo accretion rate $\dot{M}_h$ following \cite{Dekel13}, which is well understood from $\Lambda$CDM, and it is a function of halo mass and redshift. Finally, we estimated $\epsilon$ from:
\begin{equation}
    \epsilon = {\rm SFR}/f_b\dot{M}_h,
    \label{eq: SFE}
\end{equation}
\noindent where $f_b\approx0.16$ is the baryon fraction \citep{Moster18,Tacchella18}, and SFR is the derived SFR$_{10}$ from the SED modeling.
Our galaxies are hosted in halos of median (range) $\rm \log(M_h/M_\odot)=$10.75$\pm$0.01 (10.62-11.44), 10.56$\pm$0.01 (10.38-11.11), and 10.51$\pm$0.02 (10.40-10.92) at $6 < z < 7$, $7 < z < 9$, and $9 < z < 12$, respectively.
We tested the impact of a different SHMR by using the {\sc{UniverseMchine}} simulation suite \citep{Behroozi19}. We found consistent results, except at $z>9$, which is possibly due to the fact that this simulation suite was calibrated using pre-JWST observations. 

\begin{figure}[!tbp]
\includegraphics[width=1\columnwidth]{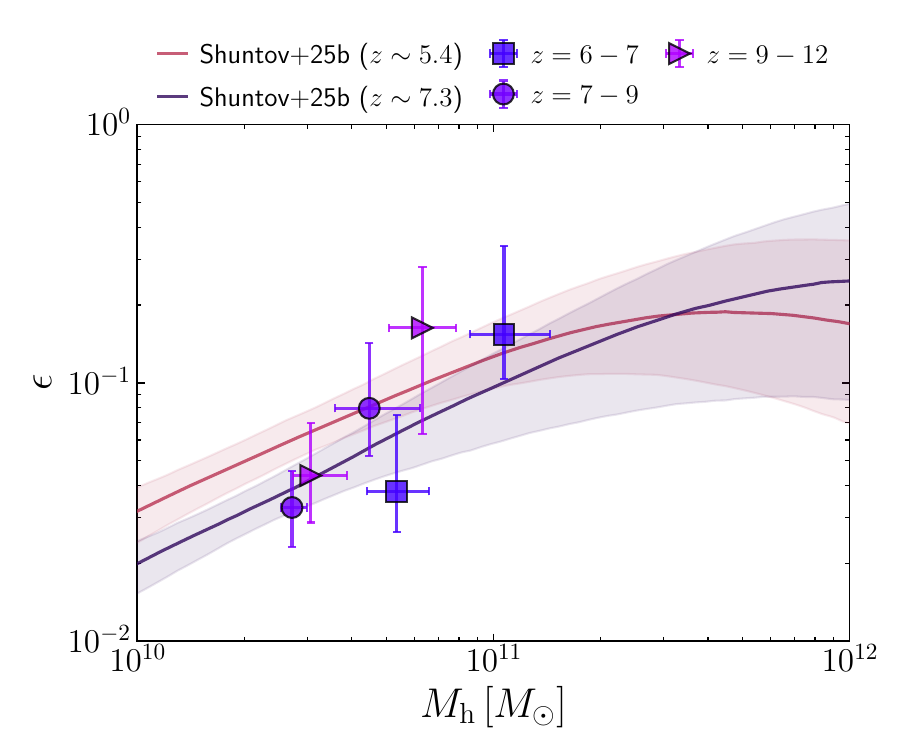}
\caption{Instantaneous star formation efficiency as a function of halo mass for the three redshift bins in our sample. Solid lines and shaded regions show the estimates from \cite{Shuntov25}, based on the UV luminosity function and halo occupation distribution (HOD) modeling. Our estimates of $\rm \epsilon$ do not show any strong variation with redshift and confirm the low values obtained by \cite{Shuntov25}.}
\label{fig: SFE}
\end{figure}

In Fig.~\ref{fig: SFE}, we show our derived $\rm \epsilon$ as a function of M$_h$ for our sample covering the low-mass end ($\log(\rm M_h/M_\odot) <$ 11.4). We do not observe any strong evolution with redshift and obtain a median $\epsilon$ of $0.06\pm0.01$, $0.06\pm0.01$, and $0.05\pm0.01$ at $6 < z < 7$, $7 < z < 9$, and $9 < z < 12$, respectively. 
We compare our results with the observational measurements from \cite{Shuntov25a}, who, using a different approach based on halo occupation distribution (HOD) and the UVLF, found a variation of $\rm \epsilon$ with halo mass consistent with our estimates. At $z<7$, our measurements agree within 0.02\,dex, while at $7 < z < 9$, our $\epsilon$ values are approximately 0.1\,dex higher. 
Our results on the SFE, shown in Fig.~\ref{fig: SFE}, do not suggest a significant (if any) increase in SFE in halos of $\log M \sim 10.3-10.8$, which combined with the measured values of $\sigma_{\rm UV}$, seems unlikely to explain the slow evolution of the observed UVLF at high-$z$ \citep[e.g.,][]{Shen23,Gelli24,Shuntov25}. However, we note that it is necessary to constrain the shape of the SFE-M$_h$ relation over a sufficiently large halo mass range (about $>2$\,dex) and constrain the position of the peak (if any) as the shape of the SFE-M$_h$ relation is more important than the peak value in the resulting UVLF \citep[e.g.,][]{Feldmann23}. Currently, our sample is too small to constrain the shape of the SFE-M$_h$ relation. Extending this work at both the fainter and brighter end will be crucial. Nonetheless, our inferred values for the SFE and $\sigma_{\rm UV}$ add further evidence that a combination with other mechanisms is likely responsible for the high-$z$ UVLF such as a negligible dust attenuation \citep[e.g.,][]{Ferrara23}, evolving top-heavy IMF \citep[e.g.,][]{Hutter25}, or locally varying SFE modulated by gas cloud density \citep[e.g.,][]{Somerville25}.

\section{Conclusion}\label{sec: conclutions}

The aim of this work is to model the SFH of high redshift galaxies ($6<z<12$) by testing different levels of stochasticity in the SFH assumption, and study the evolution of galaxies' burstiness, in order to put constraints on the UV dispersion at these redshifts.
To do so, we implemented and tested a stochastic SFH to reproduce the emission of high redshift galaxies. 

\begin{itemize}
    \item By following the approach described in \cite{CaplarTacchella19}, we modeled SFR fluctuations with time series that are computed through correlated Gaussian random numbers, defined by a PSD \citep[see also][]{Faisst24}. This approach is implemented in the SED modeling code \cigale\ and allows to well reproduce the colors of observed galaxies at $6<z<12$.
    \item Using a simulated set of galaxies' SEDs, we find that assuming a high level of stochasticity (quantified by $\sigma_{run}$) for the SFH results in better fit quality (based on $\chi^2$ distributions) than a smooth SFH assumption (close to $\tau$-delayed SFH), especially for highly bursty galaxies. This result is confirmed when applied on real observations from the JADES sample.
    \item Comparing galaxies' properties obtained for different levels of stochasticity on the SFR-stellar mass plane, we find that a smooth SFH results in a very tight relation between SFR and stellar mass at $z\geq6$ while the scatter of the relation increases with increasing level of stochasticity up to yield a weak relation at $z>7$ (from a Spearman test). This highlights again the importance of SFH assumptions in constraining galaxies MS, as already found in the literature. 
\end{itemize}

Having run \cigale\ on the JADES galaxies assuming successively different levels of burstiness, we determined the best-suited SFH for each galaxy through a Bayes factor analysis. According to this, galaxies are classified as either smooth, and their physical properties are those obtained with quasi $\tau$-delayed SFH ($\sigma_{\rm run}=0.01$), or bursty, and the results of the run using a high stochasticity level are used ($\sigma_{\rm run}=0.8$). After selecting a mass-complete sample, we find that:

\begin{itemize}
    \item For massive galaxies (8.8 $< \log M_\star/M_\odot <$ 9.5), the fraction of bursty galaxies increases from 0.38$\pm$0.08 to 0.77$\pm$0.2 with increasing redshift. This trend is not due to biases originating from differences in terms of rest-frame wavelength coverage, as using the same rest-frame coverage results in a steeper increasing trend.
    \item Although we expect low-mass galaxies to be more bursty, only $\lesssim$20\% of them are classified as bursty at all redshifts. This low fraction is attributed to observational bias, as they are faint, the S/N of the fluxes is low, and it is harder to significantly select a bursty SFH as more suitable from a Bayes factor analysis.
    \item For bursty galaxies, we find no evolution with redshift of the $\rm \log_{10}(SFR_{10}/SFR_{100})$ ratio, another indicator of burstiness. Galaxies classified as having a smooth SFH show a weak trend of decreasing ratio with decreasing redshift. Our measurements are consistent with previous estimates from the literature. However, the fraction of galaxies with $\rm \log_{10}(SFR_{10}/SFR_{100})>0.2$ increases from 0.28$\pm$0.06 to 0.38$\pm$0.11 at $z\sim6$ and $z\sim9$, respectively.
    \item We provide constraints on $\sigma_{UV}$ derived from the $M_{\rm UV}$ distribution in stellar mass bins. We obtain $\sigma_{\rm UV}$ of 0.72$\pm$0.02, 0.54$\pm$0.02, and 0.57$\pm$0.03\,mag at $6<z<7$, $7<z<9$, and $9<z<12$, respectively. These values are consistent with \cite{Shuntov25} results based on observations and the estimates of \cite{Pallottini23} and \cite{Munoz23}, confirming that bursty SFH alone is not responsible for the mild evolution of the UVLF at $z>10$.
    \item Assuming a stellar-to-halo mass relation, we use our estimated SFR to put constraints on the SFE, $\epsilon$, and obtain $0.06\pm0.01$, $0.06\pm0.01$, and $0.05\pm0.01$ at $6 < z < 7$, $7 < z < 9$, and $9 < z < 12$, respectively.
\end{itemize}

Although these results need to be confirmed with a spectroscopic sample, large enough to be statistically significant, the $\sigma_{\rm UV}$ and SFE values estimated in this work add further evidence that a combination with other mechanisms is likely responsible for the high-$z$ UVLF. Other proposed solutions, such as a negligible dust attenuation \citep[e.g.,][]{Ferrara23}, evolving top-heavy IMF \citep[e.g.,][]{Hutter25}, or locally varying SFE modulated by gas cloud density \citep[e.g.,][]{Somerville25}, must be explored.

\begin{acknowledgements}
The authors thank the anonymous referee for their valuable comments and suggestions, which helped improve the clarity and quality of this work. The authors are grateful to Guilaine Lagache for her insightful discussions and guidance. This work received support from the French government under the France 2030 investment plan, as part of the Initiative d’Excellence d’Aix-Marseille Université – A*MIDEX AMX-22-RE-AB-101.
This work was partially supported by the “PHC GERMAINE DE STAEL” programme (project number: 52217VG), funded by the French Ministry for Europe and Foreign Affairs, the French Ministry for Higher Education and Research, the Swiss Academy of Technical Sciences (SATW) and the State Secretariat for Education, Research and Innovation (SERI).
MB acknowledges support from the ANID BASAL project FB210003, supported by the French government through the France 2030 investment plan managed by the National Research Agency (ANR), as part of the Initiative of Excellence of Université Côte d’Azur under reference number ANR-15-IDEX-01.
OI and LC acknowledge the funding of the French Agence Nationale de la Recherche for the project
iMAGE (grant ANR-22-CE31-0007).
This work has received funding from the Swiss State Secretariat for Education, Research and Innovation (SERI) under contract number MB22.00072, as well as from the Swiss National Science Foundation (SNSF) through project grant 200020\_207349.

\end{acknowledgements}

\bibliographystyle{aa} 
\bibliography{bibtex} 
\newpage
\appendix
\section{Stellar mass recovery of $z>6$ galaxies}\label{app: stellar_mass}
A key difficulty in studying high redshift galaxies is the large uncertainty on the stellar mass estimation. Recent studies indicate that the lack of JWST Mid-Infrared Instrument (MIRI) photometry could lead to an overestimation of the M$_{\star}$. \cite{Williams24} found a median decrease of 0.6\,dex (up to 1 dex) in the stellar mass of red galaxies at high redshift when adding MIRI+ALMA observations, and \citep{Papovich23} found a consistent offset, $\gtrsim 0.4$\,dex, on a MIRI selected sample at high redshift.
Since our study does not include MIRI photometry, we tested any possible bias on stellar mass due to our SED fitting procedure. To do this, we added a 10\% error to the best-fitting models obtained from the first \cigale\ run on the JADES sample, and ran \cigale\ a second time to evaluate whether the same stellar masses are recovered. This approach is known as a mock test, a standard method in \cigale\ to evaluate the reliability of the derived parameters. 

\begin{figure}[!htbp]
\includegraphics[width=1\columnwidth]{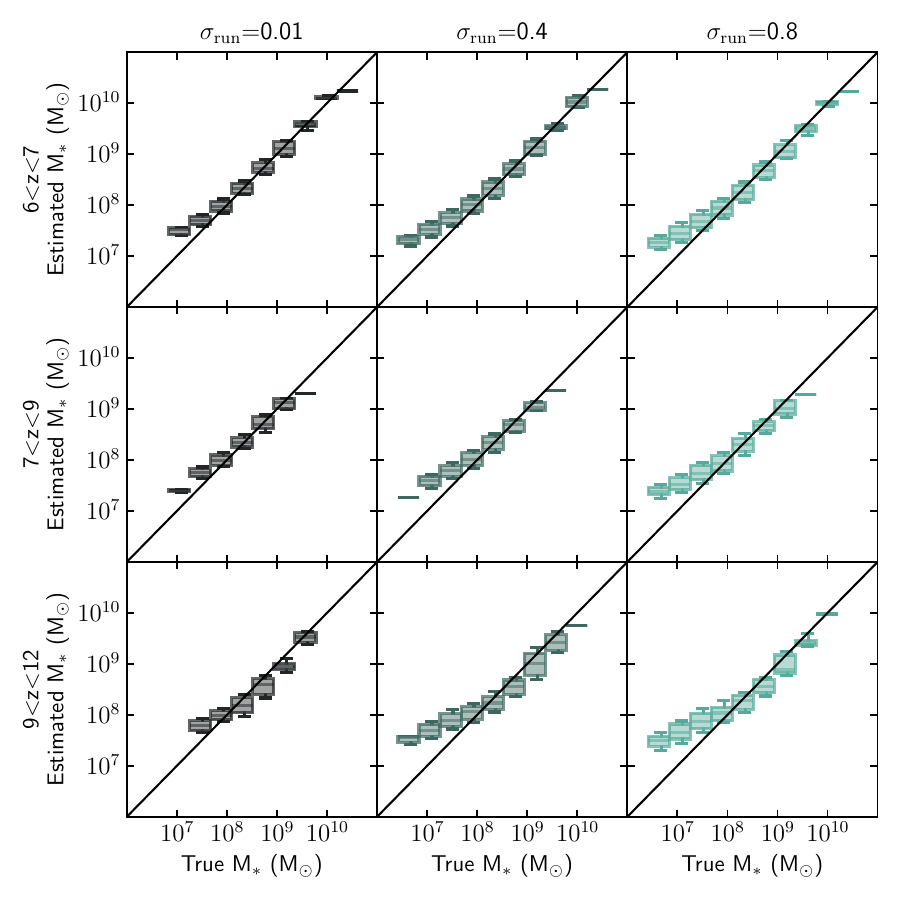}
\includegraphics[width=1\columnwidth]{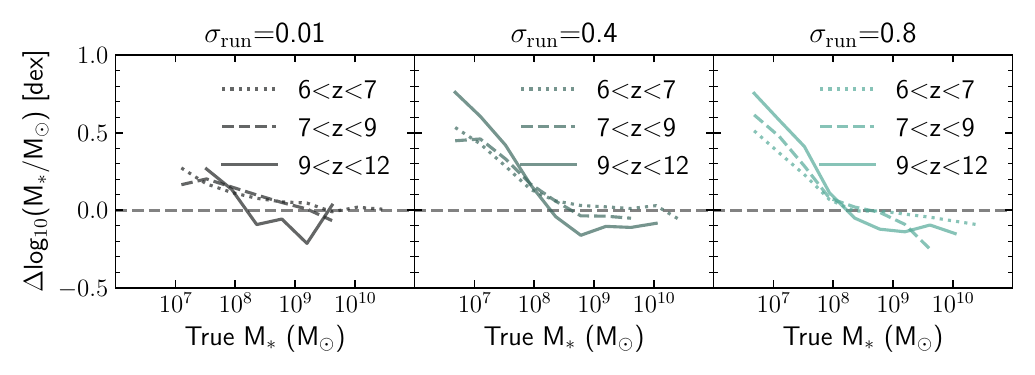}
\caption{Upper Panel: estimated stellar mass, when adding $10\%$ error to best fit and ruining {\cigale} (mock test), as function of the True stellar mass from the $\rm \sigma_{run} = 0.01$, 0.4, and 0.8 for the different redshift bins. Lower Panel: $\rm \Delta M_{\star} = \log (Estimated \ M_{\star}) - \log (True \ M_{\star}) $ as function of the True stellar mass. \\}
\label{fig: mass_mock_sigmas}
\end{figure}

We compare the true stellar masses from the best-fitting models of the first run (true M$_\star$) with the estimated stellar masses derived from the second \cigale\ run (estimated M$_\star$), for the three redshift bins considered in this work in Fig.~\ref{fig: mass_mock_sigmas}. Overall, the stellar masses are well recovered for the entire population, with an overestimation of $\rm \Delta log({M}_\star) \lesssim 0.2$\,dex for all SFH models in the different redshift bins. Additionally, the SFH model with $\sigma = 0.01$ does not reach M$_\star < 10^7$\,M$_{\odot}$ for $z>7$ galaxies. In this mass bin, which counts $\sim10,000$ galaxies, only 20 objects reach low M$_\star$: 19 at $6<z<7$ and just one galaxy at z=7.03.  

Fig.~\ref{fig: mass_mock_sigmas} reveals a stellar mass-dependent trend. At M$_\star \lesssim 10^8$\,M$_{\odot}$, the estimated stellar mass from the mock test tends to be overestimated, reaching up to $\sim0.6$\,dex at M$_\star \lesssim 10^{7}$\,M$_{\odot}$ and up to 0.8\,dex at the lower bound. This effect is slightly stronger at higher redshifts but appears independent of the SFH model. The trend may be mainly driven by the shift in the rest-frame wavelength range covered. In the intermediate mass range (M$_\star\sim 10^{8}-10^{9}$\,M$_{\odot}$), there is a small underestimation of $\lesssim 0.2$\,dex, independent of the SFH model. 
For M$_\star \gtrsim 10^9 \text{M}_\odot$, the underestimation remains $\lesssim 0.4$\,dex, although the number of galaxies in this range is low ($\sim 1.6-0.5$\,dex from low to high redshift). Interestingly, at the massive end of the stellar mass range, the trend flattens for all redshift bins. These findings are consistent with the results reported by \cite{Cochrane25}, who studied the robustness of stellar mass recovery using simulations and mock noise-free fluxes for NIRCam. 

\begin{figure}[!htbp]
\includegraphics[width=1\columnwidth]{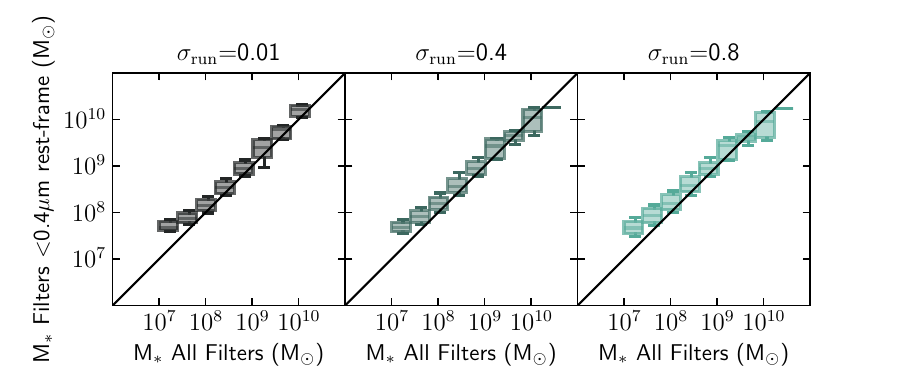}
\caption{Stellar mass derived from filters up to 0.4\,$\mu$m rest-frame as a function of the stellar mass derived from all filters, for galaxies at $6 < z < 7$ for $\rm \sigma_{run} = 0.01$, 0.4, and 0.8.}
\label{fig: mass_filter_dropped}
\end{figure}

We testd the impact of the rest-frame wavelength cover range on the estimation of M$_{\star}$ for galaxies in the last redshift bin. To account for this, we removed the longest NIRCam filters of the galaxies at $6<z<7$ to mimic the rest-frame coverage of galaxies at z$\sim11$. In Fig.~\ref{fig: mass_filter_dropped}, we compare the stellar mass distribution estimated with \cigale\ using all available filters (x-axis) and using only those covering wavelengths up to 0.4\,$\mu$m (y-axis). In an ideal scenario, we would expect a perfect match ($\rm \Delta M_\star = 0$), meaning that the rest-frame wavelength range (within JADES filters) does not affect the stellar mass estimation. However, as seen in Fig.~\ref{fig: mass_mock_sigmas}, some level of uncertainty persists because we do not cover $\gtrsim$1\,$\mu$m rest-frame. In the worst case, the impact of the rest-frame wavelength coverage (within our sample) would add to the uncertainty introduced by the absence of mid-infrared constraints. For all SFH models, the stellar masses are generally well-recovered. The mass-dependent trend is observed, with a mild overestimation that flattens around $\sim 0.2$\,dex at $\sim 10^8-10^9$\,M$_{\odot}$. At the high-mass end of the stellar mass range ($\gtrsim 10^{10}$\,M$_{\odot}$), the M$_\star$ is underestimated by $\sim 0.1$\,dex for $\sigma_{\rm run}>0.1$, though this trend is driven by a small number of sources.  

It is well known that the overestimation of M$_\star$ in the absence of rest-frame NIR constraint is primarily driven by the age-attenuation degeneracy and contamination from emission lines. Without restframe near/mid-infrared constraints, the SED fitting tends to favor younger stellar populations with higher attenuation, leading to overestimated M$_\star$ \citep{Papovich23, Williams24, Cochrane25, Wang24}.

\section{Physical properties of the sample}\label{app: physical properties}

\begin{figure*}[!htbp]
\includegraphics[width=0.24\textwidth]{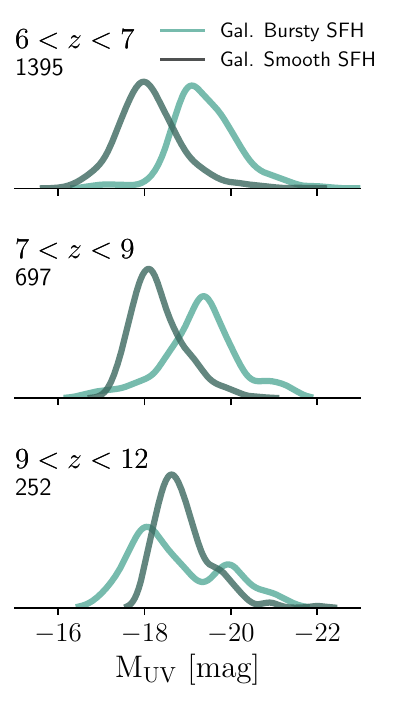}
\includegraphics[width=0.24\textwidth]{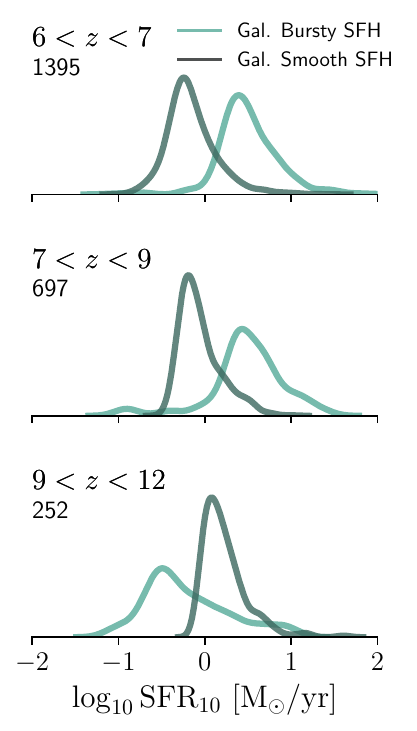}
\includegraphics[width=0.24\textwidth]{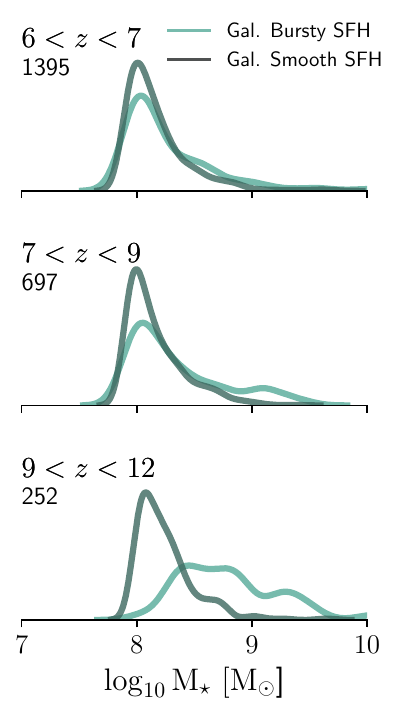}
\includegraphics[width=0.24\textwidth]{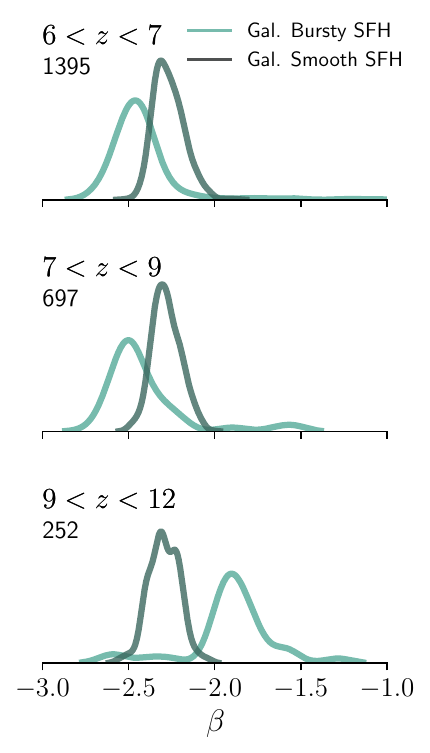}
\caption{Normalized distributions of the physical properties of classified galaxies as stochastic (light green) and $\tau-$delayed (dark green) at $6<z< 12$ in our sample. From left to right: UV magnitude, star formation rate, stellar mass and UV $\beta$ slope.} 
\label{fig: classification_prop}
\end{figure*}

We present the evolution of the physical properties of our classified sample (see Sect.~\ref{sec: galaxy_classification}) in Fig.~\ref{fig: classification_prop}. In this figure, we show the normalized distributions of UV magnitude, SFR, stellar mass, and UV slope for both galaxies with bursty and smooth SFH.
The median M$_{\rm UV}$ of bursty galaxies is brighter by 1.26 and 1.15\,mag compared to those with smooth SFHs at $6<z<7$ and $7<z<9$, respectively, while it is $\sim0.25$\,mag fainter at $z>9$. 
A similar trend is observed in the SFR distributions: bursty galaxies exhibit SFRs higher by $\sim$1\,dex at $z<9$, but lower by $\sim$0.4\,dex at $z>9$. Nevertheless, they also reach the lowest SFR values in the sample. The UV slope ($\beta$) shows a similar trend, with bursty galaxies being $\sim$0.2 bluer at $z<9$ and $\sim$0.4 redder at $z>9$. The median stellar mass shows small differences, although at $z<9$ bursty galaxies are $\sim0.5$\,dex more massive.

\section{Validation of SFH Classification with Bayes factor}\label{app: clasification_test}

\subsection{Impact of the S/N on the Classification}\label{app: clasification_test_snr}

To provide a purity estimate of our classified sample and test the impact of the S/N on the classification from the Bayes factor (see Sect.~\ref{sec: bayes_factor}), we used the best-fit models of the \cigale\ run of the JADES galaxies as simulated observations to evaluate the fraction of misclassified galaxies in each category. From this set of SEDs, we know all the properties. This simulation provides a more realistic set of data than the mock catalogs presented in Sect.~\ref{sec: module_test}, which were built from a flat grid of parameters.

For each of these SEDs, we choose either the one obtained from the fit with $\rm \sigma_{run} = 0.01$ or from the fit with $\rm \sigma_{run} = 0.8$. We thus constitute a sample of galaxies that are labeled either as ``smooth" or ``bursty." We do a new \cigale\ run on this simulated sample and apply our classification to test it, especially the impact of the S/N on it. In Table~\ref{tab: purity_test}, we provide the results of this test for several S/N bins, from 0-5 to 20-30, and for two BF thresholds. The fraction of bursty galaxies classified as bursty (true positive) increases with the S/N for both BF thresholds, and thus the fraction of bursty galaxies classified as smooth (False negative) decreases. With a BF threshold of 3, as used in our analysis, the fraction of bursty galaxies classified as such varies from 35\% to 88\% depending on the S/N. Hence, we are missing bursty galaxies, especially at low S/N (typically low-mass galaxies), but using this threshold, our sample is pure with only 5\% contamination from smooth galaxies wrongly classified as bursty. Our selection of bursty galaxies is thus quite conservative.

\begin{table*}[!htbp]
    \centering
    \caption{Validation of SFH Classification with Bayes factor for realistic simulated observations with bursty and smooth SFHs as a function of S/N.}
    \begin{tabular}{ccccccccc}
        \hline
        \hline
        
        \multirow{4}{*}{S/N bin} & \multicolumn{4}{c}{Simulated Bursty SFH} & \multicolumn{4}{c}{Simulated smooth SFH} \\
        \cline{2-9}
        & \multicolumn{2}{c}{Classify as Bursty} & \multicolumn{2}{c}{Classify as Smooth} & \multicolumn{2}{c}{Classify as Bursty} & \multicolumn{2}{c}{Classify as Smooth}\\
        & \multicolumn{2}{c}{(True positive)} & \multicolumn{2}{c}{(False negative)} & \multicolumn{2}{c}{(False negative)} & \multicolumn{2}{c}{(True positive)}\\
        & $BF>1$ & $BF>3$ & $BF>1$ & $BF>3$  & $BF>1$ & $BF>3$ & $BF>1$ & $BF>3$\\
        \hline
        \hline
        {[0--5]}    & $0.40$ & $0.13 $ & $0.60$ & $0.87$ & $0.30$ & $0.03 $ & $0.70$ & $0.97$ \\
        {[5--10]}   & $0.60$ & $0.35 $ & $0.40$ & $0.65$ & $0.31$ & $0.05 $ & $0.68$ & $0.95$\\
        {[10--20]}  & $0.76$ & $0.61 $ & $0.24$ & $0.39$ & $0.34$ & $0.05 $ & $0.65$ & $0.95$\\
        {[20--30]}  & $0.88$ & $0.88 $ & $0.22$ & $0.22$ & $0.27$ & $0.05 $ & $0.73$ & $0.95$\\
        
        \hline
        \hline
    \end{tabular}
    \tablefoot{The table shows the fraction of objects classified as bursty or smooth for each SFH type (true or false positives or negatives) using two BF thresholds ($BF>1$ and $BF>3$).}
    \label{tab: purity_test}
\end{table*}

\subsection{Performance Test}\label{app: clasification_test_fraction}

To evaluate the robustness of our classification method based on the Bayes factor (see Sect.~\ref{sec: bayes_factor}), we carried out controlled experiments using the mock catalogs built in Sect.~\ref{sec: module_test}, with the input parameters listed in Table~\ref{tab: input_param}. To reproduce realistic conditions, we added realistic noise, ensuring that the flux uncertainties follow the same distribution as in the observations. These simulated catalogs allow us to quantify possible biases in the recovered fractions and the reliability of the classification method.

In the mock catalogs we have a set of simulated observations with a given SFH model ($\rm \sigma_{mock} = 0.01,0.8$). We use \cigale\ to fit each mock catalogs using different values of $\rm \sigma_{run} = 0.01,0.8$, as for the observations (see Sect.~\ref{sec: module_test} for details). Then, we applied the classification method using BF, to evaluate the impact on the derived ``burst'' and ``smooth'' fraction. Galaxies with BF$>$3 are classified as bursty (stochastic), while those with BF$\leq$3 are considered smooth. Therefore, in a given redshift or mass bin, the fraction of bursty systems corresponds to the number of galaxies with BF$>$3 divided by the total number of galaxies.

First, we applied this classification to each individual mock catalog. In an ideal case, the recovered bursty fraction should be zero for the smooth catalog ($\rm \sigma_{mock} = 0.01$) and unity for the bursty one ($\rm \sigma_{mock} = 0.8$). The fraction of smooth galaxies misclassified as bursty remains below 15\% at all redshifts, meaning that the fraction of true smooth systems correctly classified is $\sim$85\%. Conversely, around 50\% of galaxies in the bursty catalog are correctly identified as bursty. This implies that we are underestimating the number of stochastic systems, rather than overestimating them, given the low contamination fraction from smooth galaxies.

Then, we built mock, observationally based, catalogs with different intrinsic fractions of bursty galaxies (30\%, 50\%, and 70\%) and applied the same classification approach. For the catalog with the lowest intrinsic bursty fraction (30\%), the contamination fraction of smooth galaxies is $\sim$8\%. Although this contamination appears relatively high, it is mainly driven by the large number of smooth systems in this sample, which statistically increases the number of false positives. In contrast, when stochasticity dominates  (70\%), the contamination falls to only $\sim$4\%. In this case, the purity of the smooth classification is high (around 90\%), and the underestimation of the bursty fraction becomes the main effect. In the intermediate case, with an intrinsic bursty fraction of 50\%, the contamination remains below $<$6\%.

In these tests, the computed fractions do not evolve significantly with redshift, indicating that no systematic bias is introduced that could mimic a decrease in the bursty fraction toward lower redshifts.

\end{document}